\documentclass[12pt]{article}
\usepackage{a4wide,epsfig}
\usepackage{scalefnt}

\usepackage[latin1]{inputenc}
\usepackage{amsmath}
\usepackage{amsfonts}
\usepackage{amssymb}
\usepackage[english]{babel}
\usepackage{slashed}

\newcommand{\nn}{\nonumber}
\newcommand{\be}{\begin{equation}}
\newcommand{\ee}{\end{equation}}
\newcommand{\bea}{\begin{eqnarray}}
\newcommand{\eea}{\end{eqnarray}}

\def\lQ{\Lambda_{\rm QCD}}

\newcommand{\ns}{n\!\!\!/}

\newcommand{\la}{\left\langle}
\newcommand{\ra}{\right\rangle}

\newcommand{\lp}{\left(}
\newcommand{\rp}{\right)}
\newcommand{\bc}{\begin{center}}
\newcommand{\ec}{\end{center}}
\def\epm#1#2{\hbox{${\lower1pt\hbox{$\scriptstyle +~#1$}}
\atop {\raise1pt\hbox{$\scriptstyle -~#2$}}$}}

\def\siml{{\ \lower-1.2pt\vbox{\hbox{\rlap{$<$}\lower6pt\vbox{\hbox{$\sim$}}}}\ }}

\catcode`@=11
\newcount\@tempcntc
\def\@citex[#1]#2{\if@filesw\immediate\write\@auxout{\string\citation{#2}}\fi
  \@tempcnta\z@\@tempcntb\m@ne\def\@citea{}\@cite{\@for\@citeb:=#2\do
    {\@ifundefined
       {b@\@citeb}{\@citeo\@tempcntb\m@ne\@citea\def\@citea{,}{\bf ?}\@warning
       {Citation `\@citeb' on page \thepage \space undefined}}%
    {\setbox\z@\hbox{\global\@tempcntc0\csname b@\@citeb\endcsname\relax}%
     \ifnum\@tempcntc=\z@ \@citeo\@tempcntb\m@ne
       \@citea\def\@citea{,}\hbox{\csname b@\@citeb\endcsname}%
     \else
      \advance\@tempcntb\@ne
      \ifnum\@tempcntb=\@tempcntc
      \else\advance\@tempcntb\m@ne\@citeo
      \@tempcnta\@tempcntc\@tempcntb\@tempcntc\fi\fi}}\@citeo}{#1}}
\def\@citeo{\ifnum\@tempcnta>\@tempcntb\else\@citea\def\@citea{,}%
  \ifnum\@tempcnta=\@tempcntb\the\@tempcnta\else
   {\advance\@tempcnta\@ne\ifnum\@tempcnta=\@tempcntb \else \def\@citea{--}\fi
    \advance\@tempcnta\m@ne\the\@tempcnta\@citea\the\@tempcntb}\fi\fi}
\catcode`@=12

\begin{document}

\begin{flushright}
\end{flushright}
\vspace*{1cm}
\begin{center}
  {\sc \large Deep inelastic scattering 
and factorization
\\
in the 't~Hooft Model} \\
   \vspace*{2cm} {\bf Jorge~Mondejar$^a$
and Antonio~Pineda$^b$}\\
\vspace{0.6cm}
{\it $^a$\ 
Department of Physics, University of Alberta, Edmonton, Alberta, Canada T6G 2G7
\\
[10pt]
$^b$\ Grup de F\'\i sica Te\`orica and IFAE, Universitat
Aut\`onoma de Barcelona, E-08193 Bellaterra, Barcelona, Spain\\}
  \vspace*{2.4cm}
  {\bf Abstract} \\
    \end{center}  
We study in detail deep inelastic scattering in the 't~Hooft model. 
We are able to analytically check current conservation and to obtain 
analytic expressions for the matrix elements with relative precision 
${\cal O}(1/Q^2)$ for $1-x \gg \beta^2/Q^2$. This allows us to compute 
the electron-meson differential cross section and its moments with 
$1/Q^2$ precision. For the former we find maximal violations of quark-hadron duality, 
as it is expected for a large $N_c$ analysis. For the latter we find violations of the 
operator product expansion at next-to-leading order in the 
$1/Q^2$ expansion. 
\\[2mm]

\vspace*{5mm}
\noindent

\newpage


\tableofcontents

\section{Introduction}

At its birth Quantum Chromodynamics (QCD) looked like a rather
 peculiar theory. It is constructed in terms of quarks and gluons, whereas all that one 
observes experimentally are hadrons, very specific combinations of those ``elementary" degrees of freedom. 
Indeed, when the idea of quarks and gluons was first proposed \cite{Gell-Mann:1972}, they 
were considered a mere fictitious tool to try to describe the hadron phenomenology. 
Nowadays, no one doubts their actual existence, as they leave their footprint in Deep 
Inelastic Scattering (DIS) experiments with hadrons, or in the ratio $R=\sigma_{e^+e^-\to 
\text{hadrons}}/\sigma_{e^+e^-\to \mu^+\mu-}$, for example. Nor does anyone doubt that QCD 
is the correct theory to explain their dynamics. However, thirty-six years after QCD 
was vindicated as the theory of strong interactions \cite{Gross:1973id}, we still lack a 
satisfactory analytic description of the hadrons in terms of the degrees of freedom and 
parameters that appear in its lagrangian.

The difficulty resides in the fact that, leaving aside symmetry considerations (or how symmetries 
are realized), the only quantitative and analytic computational scheme to check the dynamics of 
QCD from first principles consists in weak-coupling computations. In principle, 
those are limited to the computation of Green functions in the Deep Euclidean limit. 
The connection with experiment, however, requires the treatment of non-perturbative effects 
as well, and to relate those computations done in the Euclidean 
domain to the physical cut.

Non-perturbative effects are taken into account through perturbative factorization techniques. 
The idea behind this approach is to try to separate the non-perturbative effects 
from the perturbative ones, dividing our calculations into two pieces: one which we can calculate 
perturbatively, and another which we leave unevaluated and determine through comparison with 
experiment or lattice calculations, for example. 
Essentially, all these factorization techniques are inspired on Wilson's Operator Product 
Expansion (OPE)\cite{Wilson:1969}. The basis of the OPE is the application of the 
following relation in the deep Euclidean region\footnote{It should be mentioned that the primary 
definition of the OPE is without the time-ordering. The time-ordering introduces some ambiguities in the 
definition of the left-hand side of the equation (and consequently on the right-hand side). This is due to the 
fact that local terms in time are not fully determined (and we should also specify, 
in principle, in which frame we consider the time evolution). As a matter of principle, one may try to fix them by asking the 
correlator to have the desired transformation properties under the symmetries 
of the system. In practice, we will consider the imaginary part of the correlator and obtain the complete 
result through dispersion relations. This guarantees the desired analytic properties for the correlator.},
\be
\label{eqOPE}
i \int e^{iq x}dx T(A(x)B(0))\stackrel{q\to\infty}{\longrightarrow}\sum_n C_n^{AB}(q,\mu) O_n(\mu) ,
\ee
where $A$, $B$ are some local operators, $O_n$ are local operators with increasing dimensionality in $n$ and the 
right quantum numbers to reproduce the left-hand side, 
$C_n^{AB}$ are distributions, and $\mu$ is the renormalization scale. 
The coefficients $C_n^{AB}$ encode the physics beyond the scale $\mu$, and the local operators encode the physics below 
this scale. In QCD the coefficients are calculated using perturbation theory, and the operators are assumed to hold all 
the non-perturbative physics. This is not completely accurate, however, as perturbative effects can enter the matrix elements 
of the operators between the initial and final states, and non-perturbative effects make their way as well into the coefficients 
(for example, in the form of small-size instantons)\cite{Shifman:1985}. However, this is generally disregarded, and the OPE is 
used as a series of perturbative coefficients times some matrix elements to be determined experimentally or otherwise. The series is 
understood to be asymptotic: at some point in the expansion, non-perturbative effects are expected to cause it to break down \cite{OPE}. 
Actually, the validity of the OPE is only established in perturbation theory \cite{Zimmermann:1970}. There is no mathematical proof that 
Eq. (\ref{eqOPE}) can indeed reproduce well the (unknown) exact solution of QCD for processes 
which involve non-perturbative effects, even accepting its asymptotic nature. The validity of the OPE in these cases is just an assumption.

The connection of the OPE with the physical cut can be performed through dispersion relations. 
This is a well defined procedure, and the sum rules obtained with it are as good as the OPE 
is. But the OPE, as stated above, is also used to directly compute quantities on the physical cut.
If we knew the exact solution to QCD in the Deep Euclidean region (in a finite region) we could 
safely perform the analytic continuation from there to the physical cut, but as all we have at best are truncated expansions, this procedure can be a source of uncertainties, usually called quark-hadron duality violations 
(see \cite{Shifman:2000jv} for a general discussion).
We stress that quark-hadron duality violations are usually disregarded without a good theoretical basis. 
Typically they are only discussed, sometimes, in analysis of the vacuum polarization, and even more scarcely in 
other processes like DIS or $B$ decays like $B \rightarrow X_s \gamma$, see for instance 
\cite{Grinstein:1997xk,Blok:1997hs,Bigi:1998kc,Shifman:2000jv,Bigi:2001ys,Cata:2008ye}. Note that 
in these processes, perturbative factorization techniques, or the associated effective field theories 
like soft-collinear 
effective theory, simply neglect duality-violation effects completely. These effects can be easily seen in the large $N_c$ limit and 
quantified in the 't Hooft model (two dimensional QCD in the large $N_c$ limit \cite{hooft2}). 
We do so here for the case of DIS (see \cite{Mondejar:2006ct} for the case of $B$ decays). 

The practical version of the OPE (perturbative coefficients times non-perturbative operators \cite{OPE}) is at 
the basis of computations at large Euclidean momentum 
of (the moments in) DIS and the vacuum polarization tensor, which so far have been 
thought to be among the more solid predictions of QCD, since they are not affected by quark-hadron duality 
problems.
Therefore, the importance of setting the OPE and the factorization methods used in quantum field theories, especially 
in QCD, on solid theoretical ground can hardly be overemphasized. 
The OPE has been only partially checked in models, for instance in  
the 't~Hooft model. This theory is superrenormalizable and asymptotically free, so it is a nice ground 
on which to test the OPE\footnote{In the 't~Hooft model there are no marginal operators. Therefore, the coupling 
constant has dimensions and does not run; no renormalons should then arise.}. This was done at the lowest order in 
the OPE in Refs. \cite{callan,einhorn} for the vacuum polarization and for  DIS off a meson with nice agreement 
between the results of the model and the OPE expectations. In Ref. \cite{Mondejar:2008dt} the OPE was numerically 
checked in this model at next-to-leading order (NLO) in the $1/Q^2$ expansion, with logarithmic accuracy, for 
the vacuum polarization. In Ref. \cite{Mondejar:2008pi} the main results for DIS at NLO were presented. In particular a 
violation of the OPE was found at NLO in the $1/Q^2$ expansion. In this 
paper the details of that computation are presented. The paper is organized as follows.

In section \ref{tHooft} we review the 't~Hooft model. 
We will present the model, the semiclassical approximation to its solution \cite{einhorn,Brower:1978wm}, 
and the transition matrix elements for a vector current (in two dimensions one can also obtain from them the 
matrix elements for the axial-vector current).

In section \ref{DIS} we study DIS in the 't~Hooft model. 
We calculate the full, non-perturbative expression of the forward Compton scattering amplitude in terms of the 
't Hooft wave functions and energies. As we mentioned, we 
observe maximal duality violations in the physical cut when compared with the expression obtained from 
perturbative factorization. Analytic expressions for the matrix elements with $1/Q^2$ precision for $1-x \gg \beta^2/Q^2$ 
are also given. We then compute the forward Compton tensor and expand it in the Deep Euclidean domain with $1/Q^2$ precision. 
This result is compared to what we would obtain with the OPE. One would expect a perfect agreement at this order. However, 
surprisingly, we find that our expansion 
contains, besides the expected local matrix elements, some non-local ones at $O(1/Q^2)$, which cannot be part of 
the OPE. These non-local matrix elements arise from the constructive interference between two (non-analytic) 
oscillating terms.

In section \ref{concDIS} we present our conclusions. 
In appendix \ref{B} we present corrections 
to some formulas of Ref. \cite{Mondejar:2006ct}, where we studied duality violations in the 
context of semileptonic B decays in the 't~Hooft model with $1/m_Q^2$ precision. 
Nevertheless, the main conclusion of that paper remains unchanged. Namely, one observes 
no duality violations in the moments with $1/m_Q^2$ precision.

\section{$QCD_{1+1}$ in the large $N_c$ limit}
\label{tHooft}

The framework formed by QCD in two dimensions in the large $N_c$ limit is usually called 
the 't~Hooft model \cite{hooft2}. This model exhibits confinement: there are no free quarks, 
and the only states with finite mass are mesons (the mass of baryons grows with $N_c$), which 
are composed of exactly one quark and one antiquark, with an infinite ladder of gluons exchanged between 
them and an infinite series of ``rainbow" radiative corrections to their propagators (in the large $N_c$ 
limit only planar diagrams with no internal quark loops contribute, and in an appropriate gauge gluons don't 
interact with each other). In two space-time dimensions, the Dirac structure of the lagrangian becomes trivial and gluons can be integrated out, leaving us just with quark fields with no spinor structure. This allows us to solve the meson spectrum, which consists of an infinite tower of infinitely narrow resonances, due to the large $N_c$ limit, and features Regge behavior for large excitations. All of this makes the 't~Hooft model an attractive framework where one can exactly ``solve" QCD, and test computational techniques employed in the real world against exact results. 

In section \ref{QCD1+1} we will first present the appropriate coordinates and quantization frame to treat QCD in $1+1$ dimensions, 
and in section \ref{secthooft} we will consider its large $N_c$ limit, the 't~Hooft model; 
in section \ref{transmat} we will present the transition matrix elements for a vector current 
at leading order in $1/N_c$ (from which, in two dimensions, one can obtain the matrix elements for an axial-vector current).

\subsection{QCD$_{1+1}$ in the light front}
\label{QCD1+1}

The QCD Lagrangian is given by
\be
\label{QCDLF}
\mathcal{L}_{1+1}=-\frac{1}{4}G^a_{\mu\nu}G^{a,{\mu\nu}}+
\sum_i \bar{\psi}_i\lp i\gamma^{\mu} D_{\mu}-m_i +i\epsilon \rp \psi_i
\,,
\ee
where $D_{\mu}=\partial_{\mu}+igA_{\mu}$ and the index $i$ labels the flavour. 

One usually works with Minkowskian coordinates, and quantizes the fields 
in the equal-time frame, where fields are defined at $x^0=$ constant. 
However, in some cases a different set of coordinates and a different quantization 
frame prove to be more useful. In the so-called light-cone coordinates, 
the Dirac structure of the lagrangian becomes trivial in two dimensions, 
and once everything is expressed in these coordinates, it is natural to choose a 
quantization frame in which fields are defined at a constant value of one of the light-cone 
coordinates, and not $x^0$. This is the light-cone quantization frame \cite{Dirac:1949cp}. 
This quantization frame may be convenient when dealing with nearly massless particles. 
In four dimensions this line of research has been pursued by many groups, see \cite{Brodsky:1997de} for a review. 
In two dimensions it can be seen that it is a natural framework on which to solve QCD$_{1+1}$ in the 
large $N_c$ limit. 

\subsubsection{Light-cone coordinates}
\label{lcc}

Let us define a basis in $1+1$ dimensions with the two following light-like vectors 
(with the metric $g^{+-}=g^{-+}=2$ and zero elsewhere), 
\be
n_-^{\mu}=(1,1), \quad   n_+^{\mu}=(1,-1)
\,.
\ee
Light-cone coordinates are defined like
\be
x^+\equiv n_+\cdot x=\lp x^0+x^1\rp, 
\quad
x^-\equiv n_-\cdot x=\lp x^0-x^1\rp, \quad
\ee
which implies that
\be
x^0\equiv\frac{1}{{2}}\lp x^++x^-\rp, \quad
x^1\equiv\frac{1}{{2}}\lp x^+-x^-\rp, \quad
\ee
and
\be
\partial^-=2{\partial \over \partial x^+} 
=
{\partial \over \partial x^0}+{\partial \over \partial x^1}
=\partial_0+\partial_1 \sim p^- \,,
\partial^+=2{\partial \over \partial x^-} 
=
{\partial \over \partial x^0}-{\partial \over \partial x^1}
=\partial_0-\partial_1\sim p^+
\,,
\ee
\be
P\cdot x= \frac{P^+x^-}{2}+\frac{P^-x^+}{2}
\,,
\ee
\be
d^Dx={1\over 2}dx^+dx^-d^{D-2}x_{\bot}
\,.
\ee
For the Dirac algebra it is useful to define the corresping light-cone 
matrices
\be
\ns_+=\gamma^+, \quad \ns_-=\gamma^-
\,.
\ee
To have explicit expressions, it is useful to work with
an explicit representation of the Dirac algebra.
We will use the following Weyl-like representation for the Dirac algebra
\be
\gamma^0=\lp \begin{array}{cc} 0 & -i \\ i & 0\end{array}\rp \quad
\gamma^1=\lp \begin{array}{cc} 0 & i \\ i & 0\end{array}\rp
\,,
\ee
so that the corresponding light-cone matrices are given by
\be
\gamma^-=\gamma^0-\gamma^1=-2i\lp \begin{array}{cc} 0 & 1 \\ 0 & 0\end{array}\rp \quad
\gamma^+=\gamma^0+\gamma^1=2i\lp \begin{array}{cc} 0 & 0 \\ 1 & 0\end{array}\rp
\,.
\ee
We can define as well the following projection operators ($\gamma_5=\gamma^0\gamma^1$),
\bea
\Lambda_+&\equiv& \frac{1+\gamma_5}{2}=\frac{\gamma^0\gamma^+}{2}=
\frac{1}{4}\ns_-\ns_+=\frac{1}{4}\gamma^-\gamma^+
= \lp \begin{array}{cc} 1 & 0 \\ 0 &
  0 \end{array}\rp
\,,
\\
\nn
\Lambda_-&\equiv& \frac{1-\gamma_5}{2}=\frac{\gamma^0\gamma^-}{2}
=\frac{1}{4}\ns_+\ns_-
=\frac{1}{4}\gamma^+\gamma^-=\lp  \begin{array}{cc} 0 & 0 \\ 0 &
  1\end{array}\rp
\,.
\eea
If we write
\be
\psi= \lp \begin{array}{c} \psi_+ \\ \psi_- \end{array}\rp
\,,
\ee
then the projection operators separate the two components of the field,
\be
\Lambda_+\psi=\lp \begin{array}{c} \psi_{+} \\ 0 \end{array}\rp\ , \quad 
\Lambda_-\psi=\lp \begin{array}{c} 0 \\ \psi_- \end{array}\rp
\,.
\ee

\subsubsection{The QCD lagrangian in the light-cone frame}
\label{QCDLlc}

Once the light-cone coordinates have been defined, the next step is choosing a certain gauge, the light-cone gauge. 
In light-cone coordinates, the gluonic field is represented by the components
\be
A^{a,+}\equiv n_+\cdot A^a, \quad A^{a,-}\equiv n_-\cdot A^a \,.
\ee
The light-cone gauge consists in fixing $A^{a,+}(x)=0$; the reason for this choice will become evident in the next lines. In this gauge the QCD lagrangian in two dimensions can be written like
\be
\mathcal{L}_{1+1}=  \frac{1}{8}\lp\partial^+A^{a,-}\rp^2+\sum_i \psi_{i,+}^{\dagger}iD^-\psi_{i,+}+\psi_{i,-}^{	\dagger}i\partial^+\psi_{i,-}-m_i\lp\psi_{i,+}^{\dagger}(-i\psi_{i,-})+(-i\psi_{i,-})^{\dagger}\psi_{i,+}\rp\ ,
\ee
where $i$ is the flavour index.
Now, quantizing in the light-cone frame consists in defining the fields in this lagrangian at $x^+=$ constant. The coordinate $x^+$ plays therefore the role of time, the role of the energy being played by the 
conjugated variable $P^-$. The
other variables, $P^+$ (and $P_\perp$ in four dimensions) are kinematical. 
For instance, the $P^+_H$ component of an hadron $H$ behaves in a ``free"-particle way,
\be
P^+_{H}=\sum_i p^+_{i},
\ee
where the sum extends over all the partonic components of the bound state.
This allows one to define the variable ``$x_i$", which measures the fraction of
$P^+_{H}$ momentum carried by a given parton,
\be
x_i=\frac{p_i^+}{P_H^+} \ .
\ee

In this quantization frame the field $\psi_-$ is not dynamical (it doesn't evolve with ``time", and is therefore a constraint) and can be integrated out,
\be
\label{defp-}
\psi_{i,-}=i\frac{m_i}{i\partial^+}\psi_{i,+} \ .
\ee
 In our gauge, gluons, represented by the component $A^-$, are non-dynamical and can be integrated out as well\footnote{One should not forget that there 
is another constraint, the Gauss law, that restricts the Hilbert space of 
physical states to those which are singlet under gauge transformations. See 
for instance \cite{Gaete:1993gk}, where one can also find a quantization in the path 
integral formulation.}. 
After removing the constraints, the resulting Lagrangian can be written like
\bea
\mathcal{L}=\sum_{i}\psi^{\dagger}_{i,+}i\partial^-\psi_{i,+}  + i\sum_i\frac{m_i^2-i\epsilon}{4}\int dy^-
\psi^{\dagger}_{i,+}(x^-,x^+)\epsilon(x^--y^-)\psi_{i,+}(y^-,x^+)
\nonumber \\ +\sum_{ij}\frac{g^2}{4}
\int dy^- \psi^{\dagger}_{i,+}t^a\psi_{i,+}(x^-,x^+)|x^--y^-|\psi^{\dagger}_{j,+}t^a
\psi_{j,+}(y^-,x^+) \   ,
\label{Lagqcd11}
\eea
where we have defined
\begin{equation}
\epsilon(x)=
\left\{
\begin{array}{ll}
-1\ , & x<0 \,,\\
0 \ , & x=0 \,,\\
1 \ , & x>0 \,.
\end{array}\right.
\\
\end{equation}
Once we have the Lagrangian we can construct the Hamiltonian,
\begin{eqnarray}
\label{Pminus}
P^-=-i\sum_i\frac{m_i^2-i\epsilon}{4}\int dx^-dy^- \psi^{\dagger}_{i,+}(x^-,x^+)
\epsilon(x^--y^-)\psi_{i,+}(y^-,x^+)\nn
\\
-\sum_{ij}\frac{g^2}{4}
\int dx^-dy^- \psi^{\dagger}_{i,+}t^a\psi_{i,+}(x^-,x^+)|x^--y^-|
\psi^{\dagger}_{j,+}t^a
\psi_{j,+}(y^-,x^+) 
\ . 
\end{eqnarray}

\begin{figure}[ttt]
\begin{center}
\includegraphics[width=0.35\columnwidth]{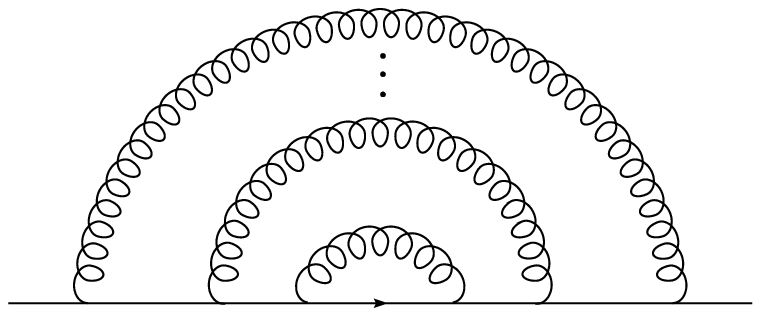}
\caption{\it The infinite series of ``rainbow" radiative corrections to the quark propagator. } 
\label{rainbow}
\end{center}
\end{figure}

The representation of the quarks in terms of free 
fields in the light-cone quantization frame is (note that this assumes that $P^2 \geq 0$)
\begin{equation}
\psi_+(x)=\int_0^{\infty}\frac{dp^+}{2(2\pi)}\left(
a(p) e^{-ipx}+
b^{\dagger}(p) e^{ipx}\right)
\,,
\end{equation}
and the anticommuting relations are
\begin{eqnarray}
&&\{a(p),a^{\dagger}(q)\}=\{b(p),b^{\dagger}(q)\}=2(2\pi) \delta(p^+-q^+) \,,\nn
 \\
&&\{a(p),b^{\dagger}(q)\}=\{b(p),a^{\dagger}(q)\}=0
\,.
\end{eqnarray}
The free propagator in the light-cone quantization frame looks like
\be
P_i^{f}\equiv\langle vac| T\lp\psi_i^f(x)\bar{\psi}_i^f(0)\rp|vac\rangle = \int \frac{d^2k}{(2\pi)^2}e^{-ik\cdot x}i\frac{k^+\frac{\gamma^-}{2}+\frac{m_i^2}{k^+}\frac{\gamma^+}{2}+m_i}{k^+k^--m_i^2+i\epsilon} \ ,
\ee
where the $f$ stands for $free$.
The renormalized propagator is given by the infinite sum of one-particle-irreducible diagrams. In the 
light-cone frame, and in the large $N_c$ limit, these diagrams are limited to the rainbow-like diagrams shown 
in Figure \ref{rainbow}, since no gluon lines can cross each other and there is no gluon 
self-interaction (the gluon lines in that diagram are not truly propagating in our gauge; strictly speaking, 
all gluon lines should begin and end at the same point in time).

However, only the first diagram of this kind contributes,
\begin{equation}
\parbox{30mm}
{
\includegraphics[width=0.25\columnwidth]{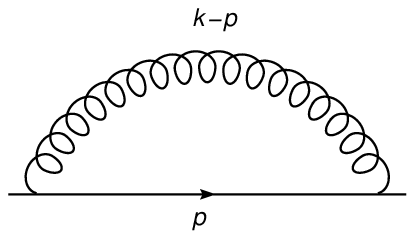}
}
\quad\quad\quad\quad\quad
=
i\beta^2\frac{\gamma^+}{2}\frac{1}{k^+} \ ,
\end{equation}
where we have defined $\beta^2 \equiv g^2 N_c/(2\pi)$.
Adding a gluon line on top of this diagram produces a vanishing integral, which kills all the other ``rainbow" diagrams. 
The infinite sum
\begin{figure}[hp]
\begin{center}
\includegraphics[width=0.9\columnwidth]{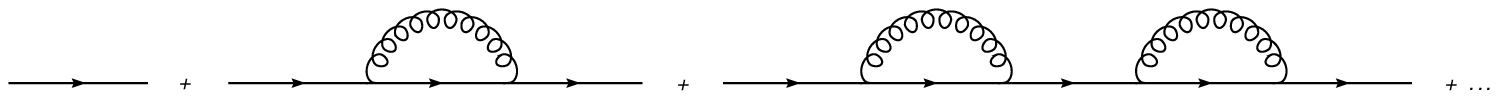}
\end{center}
\end{figure}
yields
\be
\equiv P_i(x)= \int \frac{d^2k}{(2\pi)^2}e^{-ik\cdot x}i\frac{k^+\frac{\gamma^-}{2}+\frac{m_i^2}{k^+}\frac{\gamma^+}{2}+m_i}{k^+k^--m_i^2+\beta^2+i\epsilon} \ .
\ee
Recall that this expression is gauge-dependent. Should we have chosen to quantize in the equal-time frame, 
the expression of the renormalized propagator would be (again in the large $N_c$ limit)
\be
P_i(x)^{eq. time} = \int \frac{d^2k}{(2\pi)^2}e^{-ik\cdot x}i \frac{k^+\frac{\gamma^-}{2}+\frac{k^+k^-+\beta^2}{k^+}\frac{\gamma^+}{2}+m_i}{k^+k^--m_i^2+\beta^2+i\epsilon} \ .
\ee
The difference between the two propagators is (in momentum space)
\be
P_i(p)^{eq. time}-P_i(p)^{light-cone}=i\frac{\gamma^+}{2}\frac{1}{p^+} \ ,
\ee
which illustrates the fact that the imaginary part of the propagator is independent of the quantization frame chosen. Note as well that this 
term is local in ``time", proportional to $\delta(x^+)$. Such 
terms would jeopardize the expected covariance of the Green 
function. Let us note that if, for instance, we consider the OPE of 
such Green functions at leading order in $1/Q^2$, 
we get something proportional to $\gamma^{\mu}q_{\mu}/q^2$ in the equal-time 
quantization frame but $\gamma^-/q^-$ in the light-cone quantization frame. This is not a problem by itself, since the 
propagator is not a physical quantity.

\subsection{The 't~Hooft model}
\label{secthooft}

By solving the eigenstate equation (taking into account the 
constraints, and using $n$ to schematically label the quantum numbers 
of the bound state)
\be
\label{eigenstate}
P^-|n\rangle=P_n^-|n\rangle
\,,
\ee
one obtains the basis of states over which the Hilbert space of physical 
states can be spanned. Here we will focus on the meson sector of the 
Hilbert space and we will generically label the state as $|ij;n\rangle$, 
where $i$ labels the flavour of the valence quark, $j$ labels the 
flavour of the valence antiquark and $n$ labels the excitation 
of the bound state.

The solution to Eq. (\ref{eigenstate}) in the large $N_c$ limit gives us the spectrum in the 't~Hooft model.
In this limit the sectors 
with fixed number of quarks and antiquarks are conserved and consequently 
the number of mesons; in particular, the sector with only one meson is stable in the large $N_c$ limit. Therefore, the bound state can be represented in the 
following way 
\begin{equation}
|ij;n\rangle^{(0)}=\frac{1}{\sqrt{N_c}}
\int_0^{P_n^+} \frac{dp^+}{\sqrt{2(2\pi)}}\phi^{ij}_{n}\left(\frac{p^+}{P_n^+}\right)
a_{i,\alpha}^{\dagger}(p)
b_{j,\alpha}^{\dagger}(P_{n}-p)|vac\rangle
\,,
\end{equation}
where $\alpha$ is the color index, $\phi_{n}^{ij}$ is a wave function representing the bound state, and the state is normalized as
\begin{equation}
{}^{(0)}\langle ij;m|i'j';n \rangle^{(0)}=(2\pi) 2P_n^{(0)+}\delta_{mn}\delta_{ii'}
\delta_{jj'}
\delta(P_m^{(0)+}-P_n^{(0)+})
\,.
\end{equation}
The superscript $(0)$ stands for the large $N_c$ limit, and $P_n^{(0)}$ the eigenvalue of $|ij;n\rangle^{(0)}$ (we do not explicitely 
display the flavour content of $P_n^{(0)}$ except in cases where it can 
produce confusion).

$\phi_n^{ij}$ can also be understood in terms of the gauge invariant ``null-plane" matrix element
\be
\phi_n^{ij}(x)=\frac{1}{\sqrt{N_c 2(2\pi)}}
\int dy^-e^{\frac{i}{2}y^-P_n^+x}
\langle vac|\bar{\psi}_{j,+}(0,0)\phi(0,y^-)\psi_{i,+}(y^-,0)|i,j;n\rangle^{(0)}
\ ,
\ee
where $\Phi(x^-,y^-)$ is a Wilson line,
\be
\label{Wline}
\Phi(x^-,y^-)=P[e^{(ig\int_{y^-}^{x^-}dz^-A^+(z^-))}] \ .
\ee
$P$ is a path-ordering operator. We have inserted the Wilson line between the quark fields to make gauge invariance explicit, although in the light-cone gauge ($A^+=0$) its expression is trivial.

The fact that the number of particles is quasi-conserved could make it possible 
to formulate the theory along similar lines to those of pNRQCD 
(for a review see \cite{Brambilla:2004jw}), where the wave function (the 't~Hooft 
wave function in our case) is promoted to the status of being the 
field representing the bound state. 
We will not pursue this line of research here, however.

From the large $N_c$ limit solution we can obtain the general solution to Eq. (\ref{eigenstate}) within a systematic 
expansion in $1/N_c$ using standard ``time"-independent quantum perturbation 
theory (we use $x^+$ as ``time"). It has the following structure (the momentum of the bound state 
will not be displayed explicitly unless necessary)
\bea
\label{sqrtNc}
&&
|ij;n\rangle=|ij;n\rangle^{(0)}\nn
\\
\nn
&&
+
\sum_{m,n'}\sum_k|ik;n'\rangle^{(0)}|kj;m\rangle^{(0)}{}^{(0)}
\langle ik; n'|{}^{(0)}\langle kj;m|P^-|ij;n\rangle^{(0)}
\frac{1}{P_n^{(0)-}-P_m^{(0)-}-P_{n'}^{(0)-}}
\\
&&
+O\left(\frac{1}{N_c}\right)
\,,
\eea
where the second term in the expression is $1/\sqrt{N_c}$ suppressed. 
Here we have used the fact that, at order $1/\sqrt{N_c}$, $P^-$ only 
connects neighboring sectors ($n$-mesons $\rightarrow$ $n\pm 1$-mesons), 
becoming an almost diagonal infinite dimensional matrix (see also \cite{Barbon:1994au}).

\subsection{The 't~Hooft equation}

By applying the operator $P^-$  to its eigenstate $|n\rangle$ at leading 
order in $1/N_c$
one obtains the 't~Hooft equation 
\be
\label{thoofteq}
M_n^2\phi^{ij}_n(x)={\hat P}^2\phi_n^{ij}(x)\equiv \lp \frac{m^2_{i,R}}{x}+ \frac{m^2_{j,R}}{1-x}\rp
\phi_n^{ij}(x)-\beta^2\int_0^1dy \phi^{ij}_n(y)\mathrm{P}\frac{1}{(y-x)^2}
\,,
\ee
where $M_n$ is the bound state mass, $x=p^+/P_n^+$, $p^+$ 
being the momentum of the 
quark $i$, and P stands for Cauchy's Principal Part\footnote{
One can use the following representation of this
distribution,
\begin{displaymath}
\label{pres}
P\frac{1}{(x-y)^2}=\frac{1}{2}\left[\frac{1}{(x-y+i\epsilon)^2}+\frac{1}{(x-y-i\epsilon)^2}\right]=-\frac{1}{2}\int_{-\infty}^{\infty} dz |z|
e^{i(x-y)z}
\,.
\end{displaymath}}. 
The renormalized mass is given by $m^2_{i,R}=m_i^2-\beta^2$. The
principal value prescription serves to regulate the singularity of the integrand, which originates in the infrared divergence
of the gluon propagator.

This equation cannot be solved analytically in general, but much can be said about the wave function $\phi_n(x)$ and the spectrum. The 't~Hooft wavefunctions are chosen to be real and normalized to unity
\be
\int_0^1 dx \phi_n^{ij*}(x)\phi^{ij}_m(x)=\delta_{nm}
\,,
\ee
and they vanish at the boundaries with the asymptotic
behavior
\be
\label{endpoints}
\phi^{ij}_n(x)=c^i_nx^{\beta_i}\left(1+o(x)\right)\, ,  \quad x\to 0 \ ,
\ee
and similarly for $x\to 1$ (changing $i\to j$ and $x\to 1-x$),
where $\beta_i$ is the solution of 
\be
\label{defbetai}
m_{i}^2-\beta^2+\beta^2\pi\beta_i\cot{\pi\beta_i}=0 \ ,
\ee
which in the massless limit approximates to
\be
\beta_i=\frac{\sqrt{3}}{\pi}\frac{m_i}{\beta}+o(m_i)
\,.
\ee
The only case in which the analytic solution of the 't~Hooft equation is known is the ground state with massless quarks. The solution in that case is $\phi_0(x)=1$, which means that
\be
\lim_{m_{i,j}\rightarrow 0}c^i_0=1 .
\ee 
In principle, there are several ways to 
obtain this result. One can work along the lines of Ref. \cite{hooftproc} 
to obtain an approximate Schr\"{o}dinger-like equation, which can be approximately solved 
for the ground state. Another possibility to fix the value $c^i_0$ is by matching the 
solution $\phi_0=1$ and the solution $\phi_0=c^i_0x^{\beta_i}$ in the region of overlap 
(the latter is valid for $x \ll 1$, 
whereas it can be approximated to a constant, $c^i_0$, 
for values larger than $e^{-\beta/m_i}$, which is a very small quantity for small 
masses. Therefore, there is a region on which the 
$constant$ solutions: ``$c^i_0$", and ``1", overlap and should be equal by continuity). 
One can also use the value of $\lim_{m_i\rightarrow 0} \int_0^1\phi_0(x)=1$, to fix $c^i_0$. 

The wave functions also obey the following very useful symmetry relations \cite{callan},
\be
\label{sim1}
\phi^{ij}_n(x)=(-1)^n\phi^{ji}_n(1-x) \ ,
\ee

\be
\label{sim2}
m_i\int_0^1dx\frac{\phi^{ij}_n(x)}{x}=(-1)^nm_j\int_0^1dx\frac{\phi^{ij}_n(x)}{1-x}
\,.
\ee

\subsubsection{Semiclassical solution of the 't~Hooft equation}
\label{semiclassical}

For large $n$ (high excitations) one can obtain
approximate analytic expressions both for the meson wave functions and the spectrum through a semiclassical computation.
In the interval $1/n\siml x \siml 1- 1/n$ the WKB method gives the following solution for the wave function (valid up to $O(1/n))$,
\be
\label{wf_asym}
\phi^{ij}_n(x)\simeq\sqrt{2}\,\sin[ (n+1)\pi x+\delta_n^{ij}(x)]
\,,
\ee
and the spectrum reads \cite{hooft2}
\be
\label{massn2}
M_n^2=\pi^2 \beta^2n +(m_{i,R}^2+m_{j,R}^2)\ln n+\frac{3\pi^2}{4} + 
C\lp\frac{m_{i,R}^2}{\beta^2}\rp+C\lp\frac{m_{j,R}^2}{\beta^2}\rp +O\lp\frac{1}{n}\rp \ .
\ee
The phase shift $\delta_n$ and the constant $C(m_{i,R}^2)$ in the spectrum were obtained in Ref. 
\cite{Brower:1978wm}, and they are (the expression for $C(m_{i,R}^2)$ was incorrectly written in Ref. \cite{Brower:1978wm})
\bea
\label{phaseshift}
\delta_n^{ij}(x)&=&\frac{1}{\pi}\left\{ -\frac{m_{i,R}^2}{\beta^2}[(1-x)\text{ln} n +\text{ln}x]
-(1-x)\left[C\lp\frac{m_{i,R}^2}{\beta^2}\rp-\frac{\pi^2}{8}\right]\right.\nn\\
&&\left.+\frac{m_{j,R}^2}{\beta^2}[x\text{ln}n+\text{ln}(1-x)]+x\left[C\lp\frac{m_{j,R}^2}{\beta^2}\rp
-\frac{\pi^2}{8}\right]\right\} \nn \\
C\lp\frac{m_{i,R}^2}{\beta^2}\rp&=&\frac{m_{i,R}^2}{\beta^2}\int_0^{\infty}dy\left[\frac{1-2y/\text{sinh} 2y}
{y \text{coth} y+m_{i,R}^2/\beta^2}-\frac{1}{y+\pi^2}\right]+\frac{m_{i,R}^2}{\beta^2} \ .
\eea
In the limit of small bare masses, this coefficient reads ($C(-1)=-6.07242$)
\be
C\lp\frac{m_{i,R}^2}{\beta^2}\rp=C(-1)+\sqrt{3}\pi\frac{m_i}{\beta}+{\cal O}(m_i^2)
\,.
\ee
Eq. (\ref{phaseshift}) is obtained by studying the behavior of the wave function near the classical turning points, 
which is delicate for the 't~Hooft model, and requires a precise quantum treatment of the boundary regions 
($0\le x\le 1/n$, $0\le1-x\le1/n$). This treatment is provided by the boundary-layer approximation.

\subsubsection{The boundary-layer approximation}
\label{boundary}

The boundary-layer approximation was first presented in Ref. \cite{einhorn} and later studied in Ref. \cite{Brower:1978wm}. 
As its name indicates, it is concerned with the behavior of the wave function on the boundaries, and it is valid only for large excitations.

For $x \siml 1/n$ the boundary-layer function is defined as
\be
\phi_{i}(\xi) \equiv \lim_{n \rightarrow \infty} \phi^{ij}_n\left(\xi\frac{\beta^2}{M_n^2}\right)
\,,
\ee
for finite $\xi$. For $1-x\siml 1/n$ one may use the symmetry property shown in Eq. (\ref{sim1}) to write
\be
\phi_j(\xi)=\lim_{n \rightarrow \infty}(-1)^n\phi^{ij}_n\lp1-\xi\frac{\beta^2}{M_n^2}\rp \ .
\ee
$\phi_{i}(\xi)$ approaches $\xi=0$ following the behavior of Eq. (\ref{endpoints}), and matches the WKB solution (\ref{wf_asym}) as $\xi\to \infty$,
\be
\phi_i(\xi)\simeq \sqrt{2}\, \text{sin} [(\xi/\pi+\delta^i(\xi)] \ , \quad \text{if} \quad \xi \to \infty \ ,
\ee
with
\be
\delta_i(\xi)=\frac{1}{\pi}\left\{ -\frac{m_i^2}{\beta^2}\text{ln}(\xi/\pi^2) - C\lp\frac{m_i^2}{\beta^2}\rp+\frac{\pi^2}{8}\right\} \ .
\ee

The boundary-layer function fulfills the following equation:
\be
\label{thoofteqlayer}
\phi_{i}(\xi)= \frac{m^2_{i,R}/\beta^2}{\xi}\phi_{i}(\xi)
-\int_0^\infty d \xi' \phi_{i}(\xi')
\mathrm{P}\frac{1}{(\xi'-\xi)^2}
\,.
\ee
It is possible to analytically solve the Mellin transform of this equation \cite{Brower:1978wm}. Define
\be
\psi_i(\lambda)\equiv \int_0^{\infty}d\xi \xi^{\lambda-1}\phi_i(\xi) \ .
\ee
In terms of $\psi_i(\lambda)$ the boundary-layer equation transforms into the following difference equation
\be
\label{eqdif}
\psi_i(\lambda+1)=\lp\pi\lambda \text{cot}(\pi\lambda)-\beta_i\pi \text{cot}(\pi\beta_i)\rp\psi_i(\lambda) \ .
\ee
Any solution to this equation can have an arbitrary multiplicative periodic function $P(\lambda)=P(\lambda+1)$. 
A unique solution is determined by the following three conditions: i) analyticity of $\psi(\lambda_R+i\lambda_I)$ 
in the strip $-\beta<\lambda_R<1$, which can be seen by the definition of $\psi(\lambda)$ and the behavior of 
$\phi_i(\xi)$ for small and large $\xi$; ii) real analyticity, $\psi(\lambda^*)=\psi(\lambda)$; and iii) the asymptotic behavior for $\lambda_I\to\infty$, which can be obtained from the WKB solution. With these conditions, the solution to Eq. (\ref{eqdif}) is
\be
\psi_i(\lambda)=\frac{m_i}{\beta} \psi_0(\lambda) \prod_{n=0}^{\infty} \frac{1+(m_{i,R}^2/\pi\beta^2)\text{tan}\pi\lambda/(\beta_n+n)}{1+(m_{i,R}^2/\pi\beta^2)\text{tan}\pi\lambda/(\lambda+n)} \ ,
\ee
where $\psi_0(\lambda)$ is the solution to Eq. (\ref{eqdif}) in the massless case,
\be
\psi_0(\lambda)=\pi^{\lambda}\Gamma(\lambda)\text{exp}\left[-2\pi\int_0^{\lambda-1}du \frac{u+\frac{1}{2}\text{sin}^2\pi u}{\text{sin}(2\pi u)}\right] \ ,
\ee
and $\beta_n$ are the roots $(0\leqslant \beta_n \leqslant 1)$ of
\be
\pi\beta^2(\beta_n+n)\text{cot}\pi\beta_n+m_i^2=0  \ .
\ee
In particular, one obtains for $\lambda=0,\,1$
\be
\label{boundaryint}
\int_0^{\infty} d\xi \frac{\phi_i(\xi)}{\xi}=\pi\frac{\beta}{m_i}
\,,
\qquad
\int_0^{\infty} d\xi \phi_i(\xi)=\pi \frac{m_i}{\beta}
\ .
\ee
These two expressions are the leading order contributions to the integrals $\int_0^1 dx \phi^{ij}_n(x)/x$ and $M^2_n\int_0^1dx\phi^{ij}_n(x)$, 
respectively, for large $n$. Let us see that this is indeed the case. We 
consider the first integral; if we split it into two parts, cutting at some point $\mu $ such that $1/n<\mu <1-1/n$, we can write
\bea
\label{eqmu}
\int_0^1 dx \phi^{ij}_n(x)/x&=&\int_0^{\mu} dx \phi^{ij}_n(x)/x+\int_{\mu}^1 dx \phi^{ij}_n(x)/x \nn\\
&\simeq& \int_0^{\mu} dx \phi^{ij}_n(x)/x+\int_{\mu}^1 dx \phi^{ij}_n(x) \nn \\
&=& \int_0^{\mu} dx \phi^{ij}_n(x)/x+(-1)^n\int_{0}^{\mu} dx \phi^{ji}_n(x) \nn \\
&\simeq&\int_0^{\infty} d\xi \frac{\phi_i(\xi)}{\xi}+ (-1)^n \frac{\beta^2}{M_n^2}\int_0^{\infty} d\xi \phi_j(\xi) \nn\\
&=&\int_0^{\infty} d\xi \frac{\phi_i(\xi)}{\xi} + O(1/n) \ ,
\eea
where in the third line we have used the symmetry property given in Eq. (\ref{sim1}). The integral $M^2_n\int_0^1dx\phi^{ij}_n(x)$ 
can be obtained from $\int_0^1 dx \phi^{ij}_n(x)/x$ through the 't~Hooft equation.

In Ref. \cite{Mondejar:2008dt}, by matching OPE and hadronic results for two-point correlators, 
the $1/n$ corrections to the results of Eq. (\ref{boundaryint}) were found to be
\be
\label{int1p}
\int_0^{1}dx\frac{\phi^{ij}_n(x)}{x}=\pi \frac{\beta}{m_i}
\left[
1+\frac{m_{i,R}^2+m_{j,R}^2}{2n\pi^2\beta^2}+\frac{m_im_j}{n\pi^2\beta^2}(-1)^n
+{ O}\left(\frac{1}{n^2}\right)
\right]
\,,
\ee
and
\bea
\label{int0p}
M_n^2\int_0^1dx\phi_n^{ij}(x)
&=&\pi {\beta}{m_i}
\left[
1+\frac{m_{i,R}^2+m_{j,R}^2}{2n\pi^2\beta^2}+\frac{m_im_j}{n\pi^2\beta^2}(-1)^n
\right]
\\
&&
+
(-1)^n
\pi {\beta}{m_j}
\left[
1+\frac{m_{i,R}^2+m_{j,R}^2}{2n\pi^2\beta^2}+\frac{m_im_j}{n\pi^2\beta^2}(-1)^n
\right]
+{ O}\left(\frac{1}{n^2}\right)
\,.
\nn
\eea
In Eq. (\ref{int1p}), the term $\frac{\pi\beta m_j}{n \pi^2\beta^2}(-1)^n$ 
is the contribution from the boundary $1-x\lesssim 1/n$, the term we discarded 
in the last line of Eq. (\ref{eqmu}). The other correction is therefore purely 
the $1/n$ correction to the boundary-layer function. One should obtain and solve 
the boundary-layer equation at next-to-leading order in the $1/n$ expansion to 
obtain this correction analytically. Such a computation would require a dedicated 
study and goes beyond the aim of this work.

\subsection{Transition matrix elements}
\label{transmat}

In this paper we will need the transition matrix elements between 
mesons generated by the electromagnetic interaction,
\be
\label{QED}
{\cal L}^{QED}_I
=
-\sum_ie_i\bar{\psi}_i\gamma^{\mu}A_{\mu}\psi_i
\,.
\ee
This interaction does not change flavour. Therefore, we will only consider 
neutral currents. We will consider the 
case of a charged meson made of a quark and antiquark with different flavour 
and the case of a neutral meson made of a quark and antiquark with the 
same flavour. The case of the charged meson is more interesting since 
it is stable under electromagnetic interactions. We obtain the matrix elements 
by using light-front Hamiltonian perturbation theory in the $1/N_c$ expansion, as 
we did in Ref. \cite{Mondejar:2006ct} for the case of the flavour-changing currents.
For ease of reference we review the procedure here. 

\begin{figure}
\begin{center}
\includegraphics[width=0.42\columnwidth]{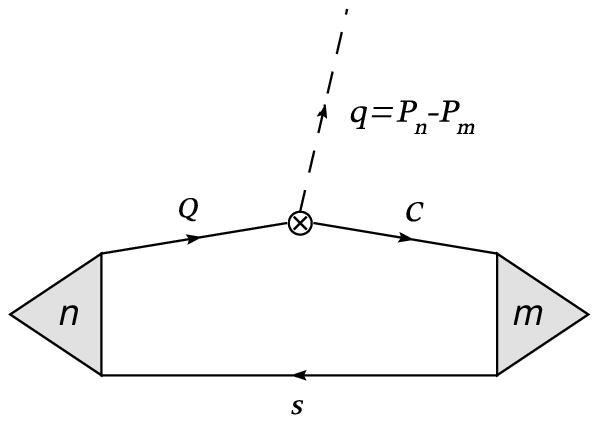}
\vskip1.5cm
\includegraphics[width=0.79\columnwidth]{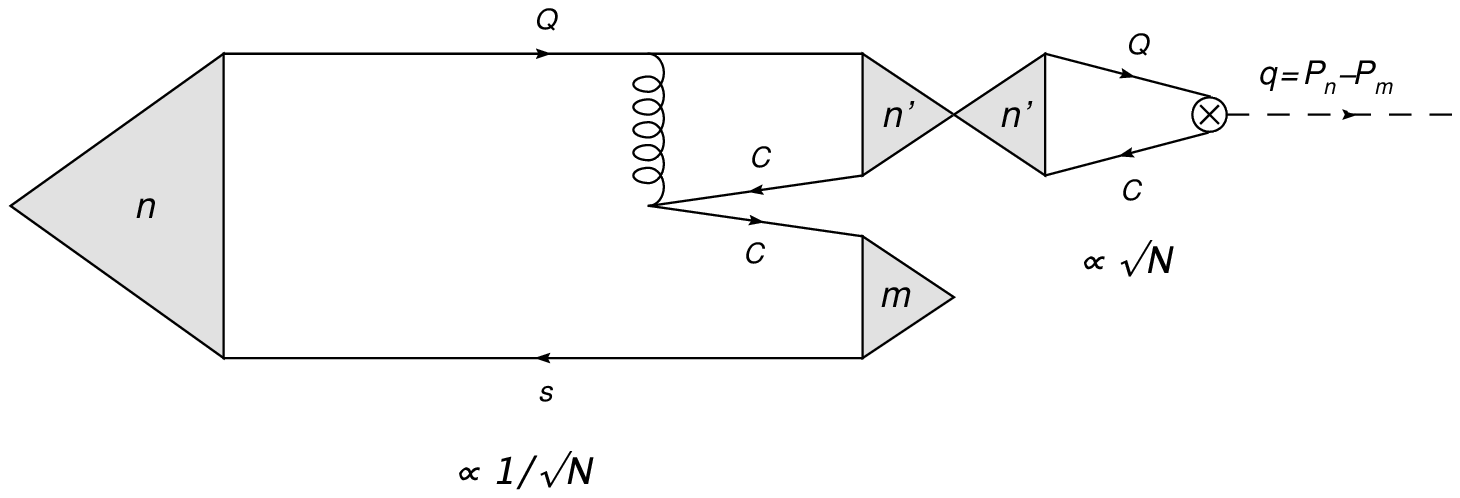}
\vskip1.5cm
\includegraphics[width=0.79\columnwidth]{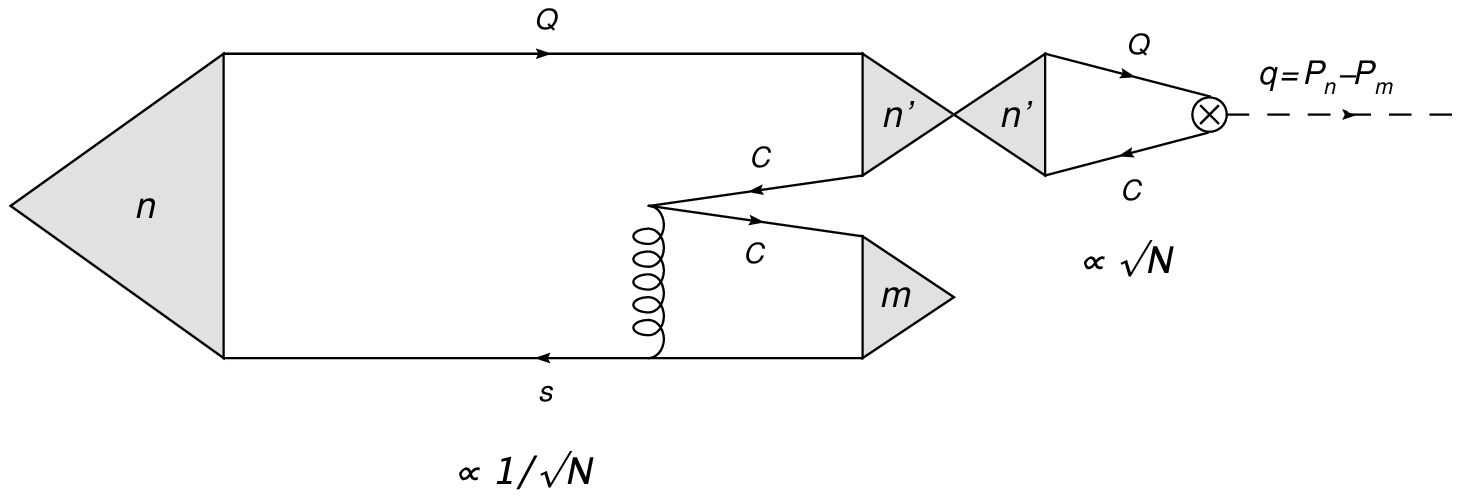}
\caption{\it Contributions to the hadronic matrix elements of the current (here shown for a general flavour-changing current
). The first 
figure corresponds to the ``diagonal" contribution to the matrix element, Eq. 
(\ref{currentdiag}). The second and third figures correspond to the 
``off-diagonal" term, Eq. (\ref{offdiag}). The $\otimes$ represents the current, 
and the gluon exchange the effective four-fermion interaction in Eq. (\ref{Pminus}).}
\label{figmatelem}
\end{center}
\end{figure}

We only aim to obtain the matrix elements at leading order in 
$1/N_c$. Nevertheless, this does not mean that we can just work with the 
leading-order solution to the bound states. As we will see, we will also need the 
$1/\sqrt{N_c}$ corrections to the bound state shown in Eq. (\ref{sqrtNc}). 

The contribution to a matrix element can be split into 
two parts. We distinguish the contributions to the current 
according to whether they come from ``diagonal" or ``off-diagonal" terms, which 
we show in Fig. \ref{figmatelem}. The ``diagonal" term directly connects the current to the 
leading $O(1/N_c^0)$ term of the bound state. Considering a flavour-changing current for some generality, the ``diagonal" term is 
\be
\label{currentdiag}
\langle i' j ;m|{\bar \psi}_{i'}\Gamma \psi_i|i j;n\rangle|_{diag.}
=
{}^{(0)}\langle i' j ;m|{\bar \psi}_{i'}\Gamma \psi_i|i j;n\rangle^{(0)}
\,.
\ee
This term is of $O(1/N_c^0)$ and is produced from terms of the 
type ${\bar \psi}_{i'}\Gamma \psi_i \sim a^{\dagger}_{i'}a_i+\cdots$, which in a way
change the flavour of the quark from $i$ to $i'$. 
Nevertheless, there is 
another possibility: ${\bar \psi}_{i'}\Gamma \psi_i \sim a_{i'}^{\dagger}b_i^{\dagger}$ ($b_{i'}a_i$), which 
can be understood as the creation (annihilation) of a new bound state. This 
possibility does not overlap with the leading-order term in the 
$1/N_c$ expansion of the bound state, but it does overlap with the $1/\sqrt{N_c}$ term. 
Whereas the matrix element connecting the one-meson sector with the two-meson sector 
is $1/\sqrt{N_c}$ suppressed, the overlap of the two-meson state with the 
current is $\sqrt{N_c}$ enhanced. This is why this contribution 
has to be considered as well at leading order in $1/N_c$.  We define the ``off-diagonal" term as
\bea
\label{offdiag}
&&\langle i' j;m|{\bar \psi}_{i'}\Gamma \psi_i|i j;n\rangle|_{off-diag.}
=
\sum_{n'}\int \frac{dP_{n'}^+}{2(2\pi)P_{n'}^+}
\frac{1}{P_n^{(0)-}-P_m^{(0)-}-P_{n'}^{(0)-}}\nn
\\
&&
\qquad
\times
\langle vac|{\bar \psi}_{i'}\Gamma \psi_i|i i';n'\rangle^{(0)}
{}^{(0)}\langle i i';n'|{}^{(0)}\langle i' j;m| P^-|i j ;n\rangle^{(0)}
\,.
\eea
This matrix element is the contribution of the term $b_{i'}a_i$. It connects the $1/\sqrt{N_c}$ correction 
to the initial state (see Eq. (\ref{sqrtNc})) with the leading final state. It is non-zero when $P_n^+\ge P_m^+$. 
A bonus of working this way is that, once 
${}^{(0)}\langle i i';n'|{}^{(0)}\langle i' j;m|P^-|i j ;n\rangle^{(0)}$ has been computed, it can be used for any current.
The other possible matrix element (involving the term $a_{i'}^{\dagger}b_i^{\dagger}$, therefore connecting the leading initial state with the $1/\sqrt{N_c}$ correction to the final state) contributes for $P_n^+\le P_m^+$. 
We will present here the matrix elements at leading order in $1/N_c$ just for the case $P_n^+ \ge P_m^+$, which then read 
\be
\langle i' j ;m|{\bar \psi}_{i'}\Gamma \psi_i|i j;n\rangle
=
\langle i' j ;m|{\bar \psi}_{i'}\Gamma \psi_i|i j;n\rangle|_{diag.}
+
\langle i' j;m|{\bar \psi}_{i'}\Gamma \psi_i|i j;n\rangle|_{off-diag.}
\,.
\ee
We can obtain the matrix elements for the case $P_n^+ \le P_m^+$ simply by exchanging the labels $n \leftrightarrow m$ in the expressions at the right-hand side of this equation. Nevertheless, we will not need them in the next sections. 

\bigskip

We present the matrix elements for a vector current, 
$\Gamma=\gamma^{\mu}$. 
From them one can also find the matrix elements for the axial-vector current, as this current can be expressed in two dimensions as a combination of the vector current and the tensor $\epsilon^{\mu\nu}$:
\be
i\gamma^5\gamma^{\mu}=\epsilon^{\mu\nu}\gamma_{\nu} \ ,
\ee
where $\epsilon^{+-}=1$, or $\epsilon^{01}=1$.

We define
\bea
q&=&P_m-P_n\\
x&\equiv&-q^+/P_n^+ \ . \nn
\eea
The expressions below are therefore valid for $x\geq 0$.

\subsubsection{Neutral currents}
\label{tmefc}

We will consider the case of a charged meson made of a quark and antiquark with different flavour 
and the case of a neutral meson made of a quark and antiquark with the same flavour.

\subsubsection{Charged meson: Non-equal mass case}

With our conventions, the full transition matrix elements for the ``+" component of the current read
\begin{itemize}
\item{$\langle ij; m|{\bar \psi}_i\gamma^+ \psi_i|ij; n\rangle
=
2\langle ij; m|{\psi}^{\dagger}_{i,+} \psi_{i,+}|ij; n\rangle
$
\bea
\label{menoteq+}
&&=
2P_n^+(1-x)
\left[
\int_0^1dz\phi^{ij}_m(z)\phi^{ij}_n(x+(1-x)z)
\right.
\\
&&
\nn
\left.
-x^2\beta^2\int_0^1\int_0^1\int_0^1dudvdz
\frac{\phi^{ij}_m(z)G_{ii}(u,v;q^2)}{(x(1-u)+(1-x)z)^2}(\phi^{ij}_n(x+(1-x)z)-\phi^{ij}_n(xu))
\right]
\eea
}
\end{itemize}
\begin{itemize}
\item{$\langle ij; m|{\bar \psi}_{j}\gamma^+ \psi_{j}|ij; n\rangle=
-2P_n^+(1-x)
\left[
\int_0^1dz\phi^{ij}_m(z)\phi^{ij}_n(z(1-x))
\right.
$
\begin{displaymath}
\left.
-x^2\beta^2\int_0^1\int_0^1\int_0^1 dudvdz
\frac{\phi^{ij}_m(z)G_{jj}(u,v;q^2)}{(1-z(1-x)-x(1-u))^2}(\phi^{ij}_n(z(1-x))-\phi^{ij}_n(1-(1-u)x))
\right]
\end{displaymath}
\be
\label{melb+}
\ee
}
\end{itemize}
where
\be
G_{ii}(u,v;q^2)\equiv  \sum_{n'=0}^{\infty}\frac{\phi^{ii}_{n'}(u)\phi^{ii}_{n'}(v)}
{q^2-M_{n'}^2}
\,.
\ee
For the ``-" component we obtain
\begin{itemize}
\item{$\langle ij; m|{\bar \psi}_i\gamma^- \psi_i|ij;n\rangle
=
2\langle ij; m|{\psi}^{\dagger}_{i,-} \psi_{i,-}|ij; n\rangle
=
2\langle ij; m|\left(\frac{m_i}{i\partial^+}\psi_{i,+}\right)^{\dagger}
\lp\frac{m_i}{i\partial^+}\psi_{i,+}\rp|ij; n\rangle
$
\bea
\label{mela-}
&&=
\frac{2}{P_n^+}\left[m_i^2\int_0^1dz\frac{\phi^{ij}_m(z)\phi^{ij}_n(x+(1-x)z)}{z(x+(1-x)z)}
+\beta^2(1-x)
\right.
\\
\nn
&&
\left.
\times
\int_0^1\int_0^1
du
dz
\frac{\phi^{ij}_m(z)\left(q^2\int_0^1dvG_{ii}(u,v;q^2)-1\right)
}{(x(1-u)+(1-x)z)^2}
(\phi^{ij}_n(x+(1-x)z)-\phi^{ij}_n(xu))\right]
\eea
}
\item{$\langle ij; m|{\bar \psi}_{j}\gamma^- \psi_{j}|ij; n\rangle=
-\text{\Large{$\frac{2}{P_n^+}$}}
\left[m_{j}^2\int_0^1dz\text{\Large{$\frac{\phi^{ij}_m(z)\phi^{ij}_n(z(1-x))}{(1-z)((1-z(1-x))}
$}}+\beta^2(1-x)\right.
$
\bea
\nn
&&
\left.
\times
\int_0^1\int_0^1
du
dz
\frac{\phi^{ij}_m(z)
\left(q^2\int_0^1dvG_{jj}(u,v;q^2)-1\right)
}{(1-z(1-x)-x(1-u))^2}
(\phi^{ij}_n(z(1-x))-\phi^{ij}_n(1-(1-u)x))\right]
\,,
\\
\label{melb-}
\eea
}
\end{itemize}
where in Eqs. (\ref{mela-}) and (\ref{melb-}) we have used that 
\be
\label{oddn}
\int dv \phi_n^{ii}(v)=\int dv \phi_n^{jj}(v)=0\ , \qquad \text{if}\; n={\rm odd}\ ,
\ee
and
\begin{equation}
\label{ident}
\int dy
\sum_{n'=0}^{\infty}M_{n'}^2\frac{\phi^{ii}_{n'}(x)\phi^{ii}_{n'}(y)}
{q^2-M_{n'}^2}
=
q^2\int dyG_{ii}(x,y;q^2)-1
\,.
\end{equation}
In order to obtain some of the above expressions we have also 
used the equations of motion in order to rewrite 
$\psi_{i,-}$ in terms of the physical component in the 
light-cone quantization frame, $\psi_{i,+}$. This is licit as
far as the current is sandwiched between physical states.

We note that the matrix elements are related by current conservation:
\be
\label{currcons}
q^+\langle ij; m|{\bar \psi}_h\gamma^- \psi_h|ij; n\rangle
=
-q^-\langle ij; m|{\bar \psi}_h\gamma^+ \psi_h|ij; n\rangle \ ,
\ee
which holds for arbitrary values of $x$ and $Q$ (and for any flavour). It is also 
useful sometimes to use $-q^-/q^+=Q^2/(x^2(P_n^+)^2)$).
Eq. (\ref{currcons}) looks quite non-trivial if we take a look to the 
explicit expressions in Eqs. (\ref{menoteq+}) and (\ref{mela-}). Nevertheless it can be shown 
to be an exact identity by a combined use of the identity Eq. (\ref{ident})
(this equality allows to rewrite the ``off-diagonal" 
term in such a way that terms with a sum over infinity intermediate states   
drop out in the difference)
and the 't~Hooft 
equation, Eq. (\ref{thoofteq}).
 
Current conservation 
also implies that the vector current matrix element can be written in the 
following way
\bea
\label{defP}
\la ij;  m | {\bar \psi}_i\gamma^{\mu} \psi_i |ij; n\ra
&=&\lp P_n^{\mu}+P_m^{ \mu}+\frac{(M_n^2-M_m^2)}{q^2}q^{\mu}\rp
P^{ij,i}_{mn}(q^2) \,,
\\
\label{defA}
\la ij;  m | {\bar \psi}_j\gamma^{\mu} \psi_j |ij; n\ra
&=&\lp P_n^{\mu}+P_m^{ \mu}+\frac{(M_n^2-M_m^2)}{q^2}q^{\mu}\rp
A^{ij,j}_{mn}(q^2) 
\ .
\eea
Obviously $P^{ij,i}_{nm}(q^2)$ and $A^{ij,j}_{nm}(q^2)$ are related by charge conjugation
symmetry:
\be
\label{relAP}
A^{ij,j}_{nm}(q^2)=-(-1)^{n+m}P^{ji,j}_{nm}(q^2)
\,.
\ee 
This property can be easily visualized using the symmetry property of the $\text{'t~Hooft}$ 
function given in Eq. (\ref{sim1}). It allows us to easily write the antiparticle currents, 
Eqs. (\ref{melb+}) and (\ref{melb-}), in terms of the particle currents, getting the correct 
$-(-1)^{n+m}$ factor.

\subsubsection{Chargeless meson: Equal mass case}

In the case where the particle and antiparticle component of the meson 
correspond to the same field, the expressions for the matrix elements 
can be simplified. By using
\bea
\la ii;  m | {\bar \psi}_i\gamma^{\mu} \psi_i |ii; n\ra
&=&\lp P_n^{\mu}+P_m^{\mu}+\frac{(M_n^2-M_m^2)}{q^2}q^{\mu}\rp
(1-(-1)^{n+m})P^{ii,i}_{mn}(q^2) 
\,,
\eea
we obtain
\begin{itemize}
\item{$\langle ii; m|{\bar \psi}_i\gamma^+ \psi_i|ii; n\rangle
=
2\langle ii; m|{\psi}^{\dagger}_{i,+} \psi_{i,+}|ii; n\rangle
$
\bea
&&
=
\left[
1-(-1)^{n+m}
\right]
2P_n^+(1-x)
\left[
\int_0^1dz\phi^{ii}_m(z)\phi^{ij}_n(x+(1-x)z)
\right.
\\
&&
\nn
\left.
-x^2\beta^2\int_0^1\int_0^1\int_0^1dudvdz
\frac{\phi^{ii}_m(z)G_{ii}(u,v;q^2)}{(x(1-u)+(1-x)z)^2}(\phi^{ii}_n(x+(1-x)z)-\phi^{ii}_n(xu))
\right]
\eea
}
\item{$\langle ii; m|{\bar \psi}_i\gamma^- \psi_i|ii;n\rangle
=
2\langle ii; m|\left(\frac{m_i}{i\partial^+}\psi_{i,+}\right)^{\dagger}
\lp\frac{m_i}{i\partial^+}\psi_{i,+}\rp|ii; n\rangle
$
\bea
\label{melii-}
&& 
=
\left[1-(-1)^{n+m}\right]
\frac{2}{P_n^+}\left[
m_i^2\int_0^1dz\frac{\phi^{ii}_m(z)\phi^{ii}_n(x+(1-x)z)}{z(x+(1-x)z)}
+\beta^2(1-x)
\right.
\\
&&
\left.
\times
\int_0^1\int_0^1
du
dz
\frac{\phi^{ii}_m(z)
\left(q^2\int_0^1dvG_{ii}(u,v;q^2)-1\right)
}{(x(1-u)+(1-x)z)^2}
(\phi^{ii}_n(x+(1-x)z)-\phi^{ii}_n(xu))
\right]\nn
\,.
\eea
}
\end{itemize}
The terms associated to ``1" correspond to the particle current, and the ones associated to ``$-(-1)^{n+m}$" to the antiparticle one.
Note that our results fulfill charge conjugation symmetry. 

The ``+" component of the vector current was already computed in Ref. \cite{einhorn}. The computation of the rest of the matrix elements had 
to wait to Ref. \cite{Burkardt:1991ea}, but we disagree with their results 
for the ``-" component of the vector current. 

\section{DIS in the 't~Hooft model}
\label{DIS}


We consider here the differential cross section of the electron-meson 
scattering going to electron+anything: $e M \rightarrow e X$.
The interaction we consider is therefore the one given in Eq. (\ref{QED}).
One should also have to include the leptons and photons, which we will not do explicitly. We will only consider the electromagnetic interaction perturbatively in $e$ at the lowest non-trivial order.

We are particularly interested in the 
situation when the momentum $q$ transferred by the virtual photon 
is very large (DIS). 
Considering DIS with mesons rather than with baryons will allow us to use the
results of the 't~Hooft model. The meson has flavour content 
$M \sim q_i\bar q_j$. Unless explicitly stated, the formulas will 
hold true either if $i=j$ or not.

DIS in QCD$_{1+1}$ could be considered somewhat delicate, since Quantum Electrodynamics (QED) is confining in two dimensions. 
However, we will consider the electromagnetic interactions as pure current insertions. Working in the 't~Hooft model we will be 
able to write down the full, non-perturbative expression of the scattering amplitude. 
As we mentioned, we 
observe maximal duality violations in the physical cut when compared with the expressions obtained from 
perturbative factorization. Analytical expressions for the matrix elements with $1/Q^2$ precision for $1-x \gg \beta^2/Q^2$ 
are also given. We then compute the forward Compton tensor in the Deep Euclidean domain with $1/Q^2$ precision and 
compare the results to what we would obtain with the OPE. Surprisingly, we find that our expansion 
contains, besides the expected local matrix elements, some non-local ones at $O(1/Q^2)$, which cannot be reproduced by 
the OPE. 

In sections \ref{kinDIS}  and \ref{scattcrosssect} we will present the kinematics and definitions that we will use, and the expression of the scattering cross section; in section \ref{matDIS} we will give the approximate form of the matrix elements with $1/Q^2$ precision; we will use these approximate matrix elements to give our result for the expansion of the forward Compton scattering amplitude in section \ref{momDIS}. Next, in section \ref{pertDIS} we will perform the perturbative calculation of the imaginary part of the amplitude; in section \ref{mompertDIS} we will obtain the OPE expansion from the previous result, and compare it with the exact expression. 

\subsection{Hadronic computation}

\subsubsection{Kinematics}
\label{kinDIS}

\begin{figure}[hhh]
\begin{center}
\includegraphics[width=0.49\columnwidth]{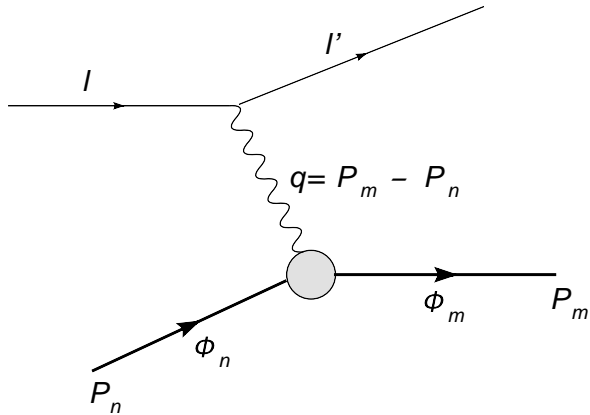}
\caption{Deep-inelastic scattering off a light meson. The 
momentum of the photon $q$ is incoming.}
\label{dis}
\end{center}
\end{figure}

The kinematics of DIS can be found in Fig. \ref{dis}. They share some similarities with those of semileptonic $B$ decays \cite{Mondejar:2006ct}. However, in B decays one has that $q^2=0$, which is an added constraint. Therefore, 
the expressions we have to deal with here are more complicated than those 
used in $B$ decays, since now more kinematical freedom is allowed. 

The kinematical variables that we use are $Q^2\equiv-q^2 >0$ (this comes from the kinematics of 
the scattering process) and ($q=P_m-P_n$)
\be
x_B\equiv\frac{Q^2}{2P_n\cdot q}\,, \qquad  x \equiv -\frac{q^+}{P_n^+}
\,.
\ee
$x_B$ is explicitly Lorentz invariant, whereas $x$ is the natural variable that 
appears in the solution of the 't~Hooft equation. Momentum 
conservation in terms of $x$ reads 
\be
\label{Pm}
P_m^2=Q^2\frac{1-x}{x}+P_n^2(1-x)
\,,
\ee
and in terms of $x_B$ reads
\be
\label{Pm2}
P_m^2=Q^2\frac{1-x_B}{x_B}+P_n^2
\,.
\ee
By combining both equations one obtains the following relation between $x$ and $x_B$
\be
\label{xeqxB}
x_B=\frac{x}{1-\frac{M_n^2}{Q^2}x^2}
\,.
\ee
This equality is quite remarkable. It means that in two dimensions 
$x$ is also Lorentz invariant. On the other hand, $x_B$ can be unambiguously obtained 
from Eq. (\ref{Pm2}):
\be
x_B=\frac{1}{1+\frac{P_m^2-P_n^2}{Q^2}} \ .
\ee
With this expression we see that if the target is in the ground state ($n=0$), 
$x_B$ is positive and runs from 0 to 1. Having $x_B>0$ means that we always have 
$q^0>0$, that is, the photon always transfers energy to the target. If the initial 
state is a resonance (stable under strong interactions in the 't~Hooft model), we 
can have $P_m^2<P_n^2$, and the value of $x_B$ is not restricted anymore. If $P_n^2<Q^2$, 
$x_B$ will run from 0 to some number greater than 1, but if $P_n^2>Q^2$, $x_B$ can reach 
negative values (and therefore so does $q^0$). We will not be concerned about this last situation 
since we will only consider the deep inelastic region, with $Q^2 \gg P_n^2$.

The fact that $x_B$ can be larger than 
1 may look surprising, compared with standard DIS in 4 dimensions. The 
reason has nothing to do with dimensions but with the fact that in four dimensions one usually 
considers the ground state as the initial state. In this situation the final state can only be the initial state or an excitation, which sets $x_B^{max}=1$,
whereas $x_B^{max} > 1$ happens for a final state with lower invariant mass than the initial state. 

If we now consider $x$, we see that Eq. (\ref{xeqxB}) is quadratic in $x$, which means that for a given $x_B$ we have two possible solutions, 
$x>0$ and $x<0$ (that is, either $q^+<0$ or $q^+>0$). This has a physical origin, the parity symmetry. 
The first solution corresponds to the frame where the initial lepton moves from right to left, 
and the second one to the frame where the lepton goes from left to right. We can give bounds to the values 
of $q^+$ and $q^-$ in both cases through momentum conservation, and the requirement that $P_m^2\geqslant 0$. The conditions are symmetric, of course,
\be
\left\{
\begin{array}{llll}
0<\frac{-q^+}{P_n^+}<1&\ ,&\frac{Q^2}{P_n^2}<\frac{q^-}{P_n^-}<\infty& \text{left-moving incoming electron}\\
&  \\
\frac{Q^2}{P_n^2}<\frac{q^+}{P_n^+}<\infty&\ ,& 0<\frac{-q^-}{P_n^-}<1 & \text{right-moving incoming electron}\ .
\end{array}\right.
\ee
However, as we break parity symmetry in the light-cone quantization frame, the two cases will not be equivalent for us. With the transition matrix elements given in sec. \ref{transmat}, it is easier to find approximate expressions for the cross section in the case of a left-moving incoming electron, and therefore this is the frame we choose. In our frame, then, $0<x_B<x_B^{max}$, where
\be
x_B^{max}= \frac{1}{1+\frac{P_0^2-P_n^2}{Q^2}} \ge 1 \ ,
\ee
and $0<x<1$, where the limit $x=1$ is only reached in the massless case when the initial and final state are in the ground state.

Note that in Eq. (\ref{Pm}), as far as $Q^2 \gg P_n^2$, 
the factor $P_n^2(1-x)$ can always be neglected (in a first 
approximation) compared with $Q^2\frac{1-x}{x}$, independently of the value of $x$. 
This is not true in Eq. (\ref{Pm2}), where in the $x_B \rightarrow 1$ limit $P_n^2$ 
can not be neglected. More generally, we can actually distinguish three different kinematical regimes:
\begin{displaymath}
\begin{array}{lllllll}
\text{a)} & P_{m}^2=M_m^2 \gtrsim Q^2 &; & m \gtrsim Q^2/\beta^2&, &1-x_B \sim 1\,,\\
\text{b)} &P_{m}^2=M_m^2 \sim  Q\lQ&; & m \sim Q/\beta&,& 1-x_B\sim \beta/Q\, , \\
\text{c)} & P_{m}^2=M_m^2\sim \lQ^2&; & m \sim 1&,& 1-x_B\lesssim \beta^2/Q^2\, .
\end{array}
\end{displaymath}
where $m$ is the principal quantum number of the hadronic excitation.

In this paper we choose to work in the target rest frame. 
The DIS limit corresponds to the limit where $Q^2\to \infty$ keeping $x_B$ fixed. 
In our frame this implies  $q^+<0$, $q^-\to \infty$ and the Bjorken $x_B$ goes like
\be
x_B\equiv\frac{-q^2}{2P_n\cdot q}\simeq -\frac{q^+}{P_n^+}\equiv x
\ee
The Bjorken variable $x_B$ and $x$ are in general different but they approach 
each other for large $Q^2$. In general the momenta in this frame scale like
\bea
P_n^+&=&P_n^-\sim \Lambda_{QCD} \sim \beta \nn\\
q^+&=& -x P_n^+ \nn\\
q^-&=&Q^2/(xP_n^+) \nn\\
P_m^+&=&P_n^+(1-x) \nn\\
P_m^-&=&P^-_n+q^-\sim q^- \ .
\eea
We have two possible expansion parameters, $\lambda \equiv \sqrt{\lQ/Q}$, which is always small, 
and $\bar \lambda \equiv \sqrt{1-x}$, which is small in the limit $x\to 1$.
We can see that in this limit the outgoing hadron behaves as a collinear 
particle, with a big momentum component $P_m^-$ and a small invariant mass, although the scaling is not standard: 
\be
P_m^+ \sim \bar \lambda^2 \lQ \ , \, P_m^- \sim Q^2/\lQ \ .
\ee

There is another frame in which DIS is usually studied, the Breit frame, in which the 
photon carries no energy. The momentum components in this frame are
\bea
q^+&=&-Q\nn\\
q^-&=&Q\nn\\
P_n^+&=&Q+l^+\nn\\
P_n^-&=&l^-\nn\\
P_m^+&=&l^+\nn\\
P_m^-&=&Q+l^- \ ,
\eea
where $l^{\pm}$ are fixed by setting $P_n^2\simeq Ql^-$ and $P_m^2\simeq Ql^+$, and
\be
x\simeq \frac{Q}{Q+l^+} \ .
\ee
In this frame the scaling is simpler: $l^+\sim Q \bar{\lambda}^2$ and $l^-\sim Q \lambda^2$, so that 
\be
P_m^+ \sim Q \bar{\lambda}^2\, , P_m^- \sim Q\ .
\ee
This is the natural frame to study DIS near $x=1$ (there is no dependence on $\Lambda_{QCD}$ neither in $P_m^+$ nor in $P_m^-$, 
unlike in the target rest frame). The price is that here both the initial and final hadron are collinear traveling  
in opposite directions. 

\subsubsection{Scattering cross section}
\label{scattcrosssect}

The differential cross section is given by
($l/l'$ represent the momentum of the incoming/outcoming lepton)

\be
\label{one}
d^2\sigma=\frac{1}{4(P_n \cdot p_l-M_n^2m_l^2)^{1/2}} \frac{dP_m^+}{2(2\pi)P_m^+} 
\frac{dl'^+}{2(2\pi)l'^+}\big|\langle P_m,l'|S|P_n,l\rangle\big|^2(2\pi)^2\delta^2(P_n +l -P_m -l')
\,,
\ee
where $S$ is the transition matrix operator, and $|l\rangle = \sqrt{l^+}a_l(l)|vac\rangle$.
Expressing $l'^{\mu}$ as $l'^{\mu}=l^{\mu}-q^{\mu}$, we can rewrite this as
\be
d^2\sigma=\frac{1}{4(P_n \cdot p_l-M_n^2m_l^2)^{1/2}} \frac{dP_m^+}{2(2\pi)P_m^+} \frac{dq^+}{2(2\pi)(l^+-q^+)}
\big|\langle P_m,l-q|S|P_n,l\rangle\big|^2(2\pi)^2\delta^2(P_n +q -P_m ) \ .
\ee
\newline
Being an observable quantity, $\sigma$ is gauge-independent, and we can choose the gauge for the 
electromagnetic field that we please. The usual choice is the gauge $A_{EM}^+=0$, but there is 
also the option $A_{EM}^-=0$. We can give compact expressions for $\sigma$ at LO in $\alpha$ in each of these gauges:
\bea
\label{sigma1}
\sigma&=&\frac{1}{4(P_n \cdot p_l-M_n^2m_l^2)^{1/2}}
\int \frac{d^2q}{(2\pi)^2}\theta(q^-)\theta(-q^+)\lp \frac{1}{(q^+)^2}\rp^2 \text{Im}[l^{++}]4\pi \overline{W}^{++} \nn\\
\eea
for the first gauge, and
\bea
\label{sigma2}
\sigma&=&\frac{1}{4(P_n \cdot p_l-M_n^2m_l^2)^{1/2}}\int \frac{d^2q}{(2\pi)^2}\theta(q^-)\theta(-q^+)\lp \frac{1}{(q^-)^2}\rp^2 \text{Im}[l^{--}]4\pi \overline{W}^{--}\nn\\
\eea
for the second one, where $\overline{W}^{++}$, $l^{++}$, etc. are the components of the leptonic and hadronic tensors, defined as
\be
l^{\mu\nu}=ie^2\int d^2x e^{-i q\cdot x} \langle l|\bar{\psi}_l(x)\gamma^{\mu}\psi_l(x)\bar{\psi}_l(0)\gamma^{\nu}\psi_l(0)|l\rangle \ ,
\ee
and
\bea
\overline{W}^{\mu\nu}&=&\frac{1}{4\pi}
\sum_{h,h'}e_he_{h'}\int d^2xe^{iqx}\langle P_n|\bar{\psi}_h(x)\gamma^{\mu}\psi_h(x)
\bar{\psi}_{h'}(0)\gamma^{\nu}\psi_{h'}(0)|P_n\rangle\nn\\
&=&\frac{1}{4\pi}
\sum_{h,h'}e_he_{h'}\sum_{m=0}^{\infty}\int \frac{dP_m^+}{2(2\pi)P_m^+} \langle P_n|\bar{\psi}_h(0)\gamma^{\mu}\psi_h(0)|P_m\rangle \langle P_m| \bar{\psi}_{h'}(0)\gamma^{\nu}\psi_{h'}(0)|P_n\rangle \nn\\
&&\times (2\pi)^2\delta^2(P_n+q-P_m) \ .
\eea
In the above sum over $h$ and $h'$, there is only a contribution when both indices are equal to $i$, $j$. 
Since in the large $N_c$ limit the spectrum is comprised of zero-width resonances, this tensor is a sum of deltas at the position of each resonance, a structure that cannot be reproduced by perturbation theory.

The leptonic tensor can be easily calculated, and at $O(\alpha^0)$ we find
\bea
\text{Im}[l^{\mu\nu}]&=&\text{Im}[ie^2\int d^2 e^{-i q\cdot x}\langle l|\bar{\psi}_l(x)\gamma^{\mu}\psi_l(x)\bar{\psi}_l(0)\gamma^{\nu}\psi_l(0)|l\rangle] \\
&=&4\pi e^2l^+\lp(l^+-q^+)\delta^{\mu +}\delta^{\nu +}+\frac{m_l^4}{(l^+)^2(l^+-q^+)}\delta^{\mu -}\delta^{\nu -}\rp\delta\lp (l^+-q^+)(l^--q^-)-m_l^2\rp \ .\nn
\eea

We can see that the imaginary part of the leptonic tensor 
obeys the identity
\be
\label{l++l--}
\frac{1}{(q^+)^2}\text{Im}[l^{++}]=\frac{1}{(q^-)^2}\text{Im}[l^{--}] \ ,
\ee
which implies
\be
\label{parity}
\frac{1}{(q^+)^2}\overline{W}^{++}=\frac{1}{(q^-)^2}\overline{W}^{--} \ ,
\ee
as required by charge conjugation symmetry.
Current conservation in 1+1 dimensions also implies that 
the hadronic green 
function can be written in terms of one single scalar function (unlike 
in four dimensions, where we have two functions for a spin-zero particle)
\be
\label{defW}
\overline{W}^{\mu\nu}(q)=
\lp P_n^{\mu}-\frac{q^{\mu}q\cdot P_n}{q^2}\rp
\lp P_n^{\mu}-\frac{q^{\mu}q\cdot P_n}{q^2}\rp
 \overline{W}(Q^2,x_B)  \ .
\ee
Therefore,
\be
\label{W}
\overline{W}=\lp \frac{2x}{q^+} \rp^2 \frac{1}{\left(1+\frac{M_n^2}{Q^2}x^2\right)^2}\overline{W}^{++}=\lp \frac{2x}{q^-} \rp^2 \frac{1}{\left(1+\frac{M_n^2}{Q^2}x^2\right)^2}\overline{W}^{--} \ .
\ee

If one writes the hadronic form factor in term of the current 
matrix elements one obtains
\be
\label{structure}
\overline{W}(Q^2,x_B)=\frac{1}{2}
\sum_{m=0}^{\infty}
|e_iP_{nm}^{ij,i}(q^2)+e_jA_{nm}^{ij,j}(q^2)|^2\delta((P_n+q)^2-P_m^2)
,
\ee
where $P_{nm}^{ij,i}$ and $A_{nm}^{ij,j}$ were defined in Eqs. (\ref{defP}) and (\ref{defA}).

The delta of momentum conservation implies that $\overline{W}(Q^2,x_B)\ne 0$ only for $x_B>0$ (because we set $P_n^2<Q^2$).
We could define different hadronic 
tensors that would give the same result for the cross section, but would also have support for negative values of $x_B$. Instead of defining $\overline{W}^{\mu\nu}$ with $J^{\mu}(x)J^{\nu}(0)$ we could also define $W^{\mu\nu}$ using $[J^{\mu}(x),J^{\nu}(0)]$, 
or $\widetilde{W}^{\mu\nu}$ with $\{J^{\mu}(x),J^{\nu}(0)\}$. The scalar part of these alternative tensors would be
\bea
W(Q^2,x_B)&=&\overline{W}(Q^2,x_B)-\overline{W}(Q^2,-x_B) \\
\widetilde{W}(Q^2,x_B)&=&\overline{W}(Q^2,x_B)+\overline{W}(Q^2,-x_B) \ .
\label{Wtilde}
\eea
Note that $W(Q^2,x_B)=-W(Q^2,-x_B)$ and $\widetilde{W}(Q^2,x_B)=\widetilde{W}(Q^2,-x_B)$. 
Actually, the tensor $W^{\mu\nu}$ is often used to study DIS.

\subsubsection{Matrix elements in the DIS limit}
\label{matDIS}

In section  \ref{tmefc} we showed the matrix elements for flavour-neutral currents.
Our aim here is to obtain analytic expressions for these matrix elements 
with relative precision ${\cal O}(1/Q^2)$ in the situation when $P_m^2 \gg \beta^2$ 
(this means $1-x \gtrsim \beta/Q$, and large $m$). 
In this situation we can use the boundary-layer function and its properties for the final state 
$m$.

We first consider the ``$\gamma^-$" current, for the more general case of a charged meson ($i\ne j$). 
We will show only the particle matrix elements, as the antiparticle elements can be obtained from them using Eq. (\ref{relAP}). 
The ``diagonal" term is 
\be
\label{eqdiagonal}
\langle ij;m| \bar{\psi}_i\gamma^-\psi_i|ij;n\rangle|_{diag}=\frac{2m_i^2}{P_n^+}\int_0^1dz\frac{\phi^{ij}_m(z)\phi^{ij}_n(x+(1-x)z)}{z(x+(1-x)z)} \ .
\ee
We are only interested to compute these matrix elements for values of $x=x_m$ satisfying 
\be
P_m^2=Q^2\frac{1-x_m}{x_m} + P_n^2(1-x_m)  \gg \beta^2\ ,
\ee
since this is the requirement imposed by the delta of momentum conservation of $\bar W^{\mu\nu}$. In this situation we can assume that 
$P_m^2\simeq Q^2\frac{1-x_m}{x_m}$ and ($z\equiv \xi\, \beta^2/M_m^2$)
\be
z\frac{1-x_m}{x_m}\to \xi\frac{\beta^2}{M_m^2}\frac{1-x_m}{x_m}\simeq \xi\frac{\beta^2}{Q^2}\ll 1 \ .
\ee
Note however that this had not been true if we had kept $m$ fixed but independent of $x$ and we had performed the 
limit $x\rightarrow 0$. 

We can then expand Eq. (\ref{eqdiagonal}) by considering that the wave function $\phi_m^{ij}(z)$ oscillates heavily except 
in a small region around the origin. In a boundary-layer-like fashion, we could say that 
the integrand is concentrated between 0 and some finite $\xi\ll M_m^2/\beta^2$ and we can expand in $z(1-x)/x\ll 1$,
\bea
\label{eqlayer}
&&
\langle ij;m| \bar{\psi}_i\gamma^-\psi_i|ij;n\rangle|_{diag}\simeq
\frac{2m_i^2}{P_n^+}
\left(
\frac{\phi_n^{ij}(x)}{x}\int_0^{1}\frac{dz}{z}{\phi}_{m}^{ij}(z)+{\phi'}_{n}^{ij}(x)\frac{1-x}{x}\int_0^{1}dz {\phi}_{m}^{ij}(z)
\right.\nn
\\
\nn
&&
\left.
\qquad\qquad\qquad\qquad\qquad
-\frac{\phi_{n}^{ij}(x)}{x^2}(1-x)\int_0^{1}dz{\phi}_{m}^{ij}(z)
\right)
+\cdots
\\
\nn
&&=
2\pi\beta\frac{m_i}{-q^+}\left[ \lp 1+\frac{m_{i,R}^2+m_{j,R}^2}{2m\pi^2\beta^2}+\frac{m_im_j}{m\pi^2\beta^2}(-1)^m-\frac{m_i^2+(-1)^mm_im_j}{Q^2}\rp
\phi^{ij}_n(x)\right.\nn\\
&&\left.+\frac{(m_i^2+(-1)^mm_im_j)}{Q^2}x{\phi'}^{ij}_n(x)\right]
+o\left(\frac{1}{Q^2}\right)\ . 
\eea
In the last two terms we have used that $M_m^2\simeq Q^2\frac{1-x}{x}$. 

\begin{figure}[hhh]
\begin{center}
\includegraphics[width=0.75\columnwidth]{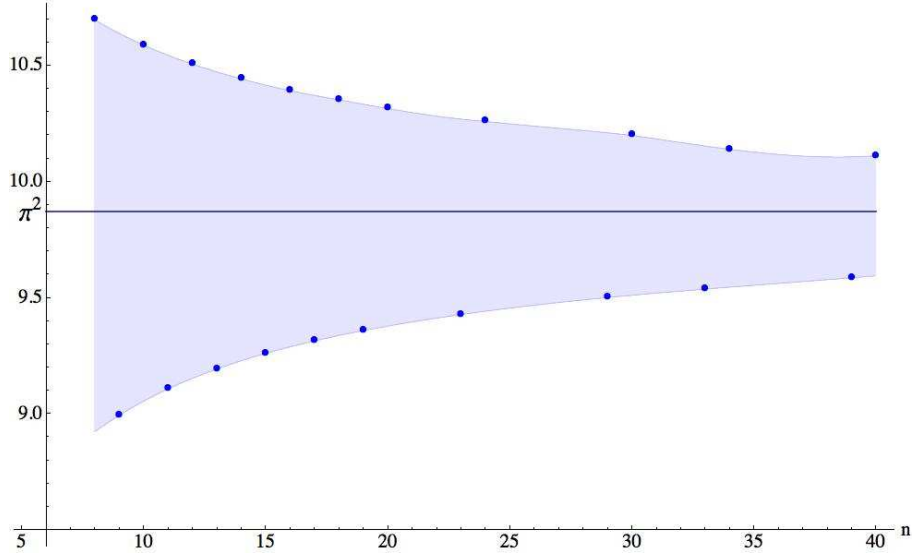}
\caption{\it Numerical evaluation of Eq. (\ref{Integral_DIS}) ranging from $n=8$ up to $n=40$. It oscillates around the expected value 
with increasing accuracy for increasing $n$ until the numerical accuracy of the computation deteriorates. 
Calculations are done with $m_i=m_j=\beta$.} 
\label{offdapp}
\end{center}
\end{figure}

The ``off-diagonal" term is
\bea
\label{offdDIS}
&&\langle ij;n| \bar{\psi}_i\gamma^-\psi_i|ij;n\rangle|_{off-diag}=\frac{2}{P_n^+}
\beta^2(1-x)
\sum_{n'=0}^{\infty}\frac{M_{n'}^2}{q^2-M_{n'}^2}\\
&&\times\int_0^1\int_0^1\int_0^1dudvdz
\frac{\phi^{ij}_m(z)\phi^{ii}_{n'}(u)\phi^{ii}_{n'}(v)}{(x(1-u)+(1-x)z)^2}(\phi^{ij}_n(x+(1-x)z)-\phi^{ij}_n(xu))  \ .\nn
\eea
The main contributions to the integral over $z$ must come from the end points of the function $\phi^{ij}_m(z)$, as $m$ is large and the wave function oscillates heavily anywhere else. We then focus on the regions
\be
z= \xi\frac{\beta^2}{M_m^2} \ ,\qquad z=1-\xi\frac{\beta^2}{M_m^2} \ ,
\ee
with finite $\xi$. The region $z=1-\xi \beta^2/M_m^2$ is actually subleading. Then, considering that, in the sum over $n'$, states with large $n'$ are weighted more than those with a small $n'$, we use the boundary layer for $\phi_{n'}(u)$ and $\phi_{n'}(v)$ as well. We use Eq. (\ref{ident}) to express the sum in terms of the Green function. Defining
\be
 u=\eta\frac{\beta^2}{Q^2} \, , \quad v=\nu\frac{\beta^2}{Q^2} \ ,
  \ee
  
it can be shown that \cite{einhorn}
\be
\lim_{Q^2\to\infty} \sum_{n'=0}^{\infty}
\frac{M_{n'}^2}{q^2-M_{n'}^2}\int_0^1dv\phi^{ii}_{n'}\lp \eta\beta^2/Q^2\rp\phi^{ii}_{n'}(v)
=h^{i}_{-}(\eta)-1 \simeq -\frac{m_i}{\pi\beta}\int_0^{\infty}d\nu \frac{\phi_i(\nu)}
{\nu+\eta+i\epsilon} \, , 
\ee
where
\be
h^{i}_{-}(\eta)\equiv \lim_{Q^2\to\infty} q^2\int_0^1 dv G_{ii}(\eta\beta^2/Q^2, v, q^2) 
\,.
\ee  
Combining all this information we can approximate the ``off-diagonal" term by
\bea
\label{offdg}
\langle ij;n| \bar{\psi}_i\gamma^-\psi_i|ij;n\rangle_{off-diag}&\simeq& -2 \frac{\beta^2}{-q^+}\frac{\beta^2}{Q^2}x\phi_n'^{ij}(x)\frac{m}{\pi\beta}\int_0^{\infty} d\xi \int_0^{\infty} d\eta \int_0^{\infty} d\nu \frac{\phi_i(\xi)\phi_i(\nu)}{(\eta+\xi)(\eta+\nu)}\nn\\
&\simeq&-2\pi\beta\frac{m_i}{-q^+}
\frac{\beta^2}{Q^2}x\phi_n'^{ij}(x)
\,,
\eea
up to $o\left(1/Q^2\right)$ terms. The last equality is found assuming that the integral is dominated by the region $\xi$,$\nu\to\infty$, where we can approximate the behavior of the boundary-layer functions by $\phi_{i}(\xi)\stackrel{\xi\to\infty}{\longrightarrow} \sqrt{2}\, \text{sin} (\xi/\pi)$:
\be
\int_0^{\infty} d\xi \int_0^{\infty} d\eta \int_0^{\infty} d\nu \frac{2 \text{sin}(\xi/\pi)\text{sin}(\nu/\pi)}{(\eta+\xi)(\eta+\nu)}=\pi^2 \ .
\ee
We have tried to confirm this result by numerically computing the integral
\be
\label{Integral_DIS}
\frac{M_{n'}^2}{\beta^2}\int_0^1 dx \int_0^1 dy \int_0^1 dz \frac{\phi_{n'}^{ij}(x)\phi_{n'}^{ij}(y)}{(z+x)(z+y)} \ ,
\ee
which, for large $n'$, tends to the integral in Eq. (\ref{offdg}). 
Unfortunately, the numerical 
computation of this integral is delicate (we use the method developed in \cite{Brower:1978wm} for our numerical calculations), 
as it requires a fine tuning between 
a small value of the integral and the large value of $M_n^2$ (for instance for $n=40$ one 
has $M_n^2/\beta^2=402.2$). Moreover, for large values of $n$ the integral becomes less precise. 
Nevertheless, the results we obtain appear to approximately converge 
(in an oscillating way) to the expected value, $\pi^2$, as can 
be seen in Fig. \ref{offdapp}.

Adding up the ``diagonal" and the ``off-diagonal" approximations, the total result for the ``-" current reads  (for $x=x_m$),
\bea
\label{-DIS}
&&\langle ij; m|{\bar \psi}_i\gamma^- \psi_i|ij; n\rangle =
2\pi\beta\frac{m_i}{-q^+}\nn\\
&&\times\left[ \lp 1+\frac{m_{i,R}^2+m_{j,R}^2}{2m\pi^2\beta^2}+\frac{m_im_j}{m\pi^2\beta^2}(-1)^m-\frac{m_i^2+(-1)^mm_im_j}{Q^2}\rp
\phi^{ij}_n(x_m)\right.\nn\\
&&\left.+(m_{i,R}^2+(-1)^mm_im_j)\frac{x_m}{Q^2}\phi_n'^{ij}(x_m)\right]
+o\left(\frac{1}{Q^2}\right)\ . 
\label{g-DIS}
\eea

For the ``$+$" current, Eq. (\ref{menoteq+}), we cannot proceed in a similar way since both the 
``diagonal" and ``off-diagonal" terms are logarithmically divergent at 
$O(1/Q^4)$. The ``diagonal" term in this case is $O(1/Q^2)$, whereas 
the ``off-diagonal" term is $O(1/Q^4)$. It is possible 
to give an approximate expression for the diagonal term at the lowest non-trivial 
order
\be
\label{+b}
2P_n^+(1-x_m)
\int_0^1dz\phi_m^{ij}(z)\phi_n^{ij}(x_m+(1-x_m)z)
=
2x_mP_n^+\pi\beta\frac{m_i}{Q^2}\phi^{ij}_n(x_m)+o\left(\frac{1}{Q^2}\right) \ ,
\ee
In order to reach the desired ${\cal O}(1/Q^4)$ accuracy, we rewrite the ``+" 
matrix element such that the logarithmic behavior cancels (or in other words 
we rewrite it in terms of the ``-" current using current conservation):
\be
\label{g+DIS}
\langle ij; m|{\bar \psi}_i\gamma^+ \psi_i|ij; n\rangle=
x^2\frac{(P_n^+)^2}{Q^2}\langle ij; m|{\bar \psi}_i\gamma^- \psi_i|ij; n\rangle \ ,
\ee
which holds both for particle and antiparticle.

Summarizing, we have obtained simplified analytic expressions for the 
``-" and ``+" particle currents with relative accuracy $O(1/Q^2)$ in the situation 
$1-x \gg \beta^2/Q^2$. For both currents we can see that the off-diagonal term is 
a correction compared with the diagonal term. The antiparticle matrix element can 
be obtained from symmetry arguments. 
\bigskip

In order to have a complete control over the matrix element, we need an estimate 
for the region $1-x \sim \beta^2/Q^2$ as well. When $x\to 1$ we can approximate $x \simeq 1- M_m^2/Q^2$, and express the matrix element for the ``$+$" current as  \cite{einhorn}
\bea
\label{1x}
&&\langle ij; m|{\bar \psi}_i\gamma^+ \psi_i|ij; n\rangle|_{x\simeq 1-M_m^2/Q^2, Q^2\rightarrow \infty}
=
2P_n^+
c^j_n\left(\frac{M_m^2}{Q^2}\right)^{1+\beta_j}
\left[
\int dz\phi^{ij}_m(z)(1-z)^{\beta_j}
\right.\nn
\\
&&\left.-
\frac{1}{M_m^2}\int_0^1dz\phi^{ij}_m(z)\int_0^{\infty}dv\frac{(1+v)^{\beta_j}-(1-z)^{\beta_j}}{(v+z)^2}h^i_{-}(M_m^2v)
\right]
+o\left(\frac{1}{Q^{2+2\beta_j}}\right)
\ ,\nn
\\
\eea
where $\phi^{ij}_n(x)=c^j_n(1-x)^{\beta_j}+o((1-x)^{\beta_j})$. The matrix element in this limit
is suppressed by a relative factor $1/Q^{2\beta_j}$ with respect to the leading term in 
Eq. (\ref{g+DIS}) (though it is enhanced with respect to the subleading term). 
Nevertheless, Eqs. (\ref{g+DIS}) and (\ref{1x}) cannot truly be compared since 
they refer to different regions in $x$. In any case, 
we will see that the associated contribution to the moments is subleading 
(as far as $N$ is not very large).
This is due to the fact that it only contributes in a narrow portion of the 
total integral of the moment. 

In principle, Eq. (\ref{g+DIS}) and (\ref{1x}) (at least the leading order expression) should merge 
in the intermediate region when $1-x$ is small and yet $M_m^2 \gg \beta^2$. We can see that they do if we let $M_m^2$ increase in Eq. (\ref{1x}).

\subsubsection{The hadronic tensor}

The expression for the component of the hadronic tensor $\overline{W}^{--}$ reads
\bea
\nn
\overline{W}^{--}&=&\frac{1}{4\pi}
\sum_m\int\frac{dP_m^+}{2(2\pi)P_m^+}
\left|\langle ij; m|\sum_{h}e_h\bar{\psi}_h(0)\gamma^-\psi_h(0)|ij;n\rangle\right|^2(2\pi)^2\delta^2(q+P_n-P_m)
\\
&=&
\frac{1}{2}
\sum_m
\left|\langle ij; m|\sum_{h}e_h\bar{\psi}_h(0)\gamma^-\psi_h(0)|ij; n\rangle\right|^2 
\delta\lp M_m^2-M_n^2(1-x)-Q^2\frac{1-x}{x}\rp 
\ .\nn
\\
\label{+1}
\eea
Using Eq. (\ref{-DIS}) we can give an approximate expression for $\overline{W}^{--}$, valid for large $Q^2$ and $1-x \gg \beta^2/Q^2$,
\bea
\overline{W}^{--}
&\simeq&
2\left(\frac{\pi\beta}{q^+}\right)^2
\sum_m \delta\lp M_m^2-M_n^2(1-x)-Q^2\frac{1-x}{x}\rp\nn 
\\
\nn
&&
\times
\left\{
e_im_i \left[ \lp 1+\frac{m_{i,R}^2+m_{j,R}^2}{2m\pi^2\beta^2}+\frac{m_im_j}{m\pi^2\beta^2}(-1)^m-\frac{m_i^2+(-1)^mm_im_j}{Q^2}\rp
\phi^{ij}_n(x)\right.\right.\nn\\
&&\left.+ (m_{i,R}^2+(-1)^mm_im_j)\frac{x}{Q^2}\phi_n'^{ij}(x)\right]\nn
\\
\nn
&&
\left.
-(-1)^{m}
		 e_jm_j \left[ \lp 1+\frac{m_{j,R}^2+m_{i,R}^2}{2m\pi^2\beta^2}+\frac{m_jm_i}{m\pi^2\beta^2}(-1)^m-\frac{m_j^2+(-1)^mm_jm_i}{Q^2}\rp
\phi^{ij}_n(1-x)\right.\right.\nn\\
\label{+hadr_lead}
&&\left.\left.-(m_{j,R}^2+(-1)^mm_jm_i)\frac{x}{Q^2}\phi_n'^{ij}(1-x)\right]
\right\}^2
\ ,
\eea
at $O(1/Q^2)$, and consequently $\overline{W}$ reads
\bea
\overline{W}
&\simeq&
2\left(\frac{2\pi\beta x}{Q^2}\right)^2
 \frac{1}{\left(1+\frac{M_n^2}{Q^2}x^2\right)^2}
\sum_m \delta\lp M_m^2-M_n^2(1-x)-Q^2\frac{1-x}{x}\rp \nn
\\
&&
\times
\left\{
e_im_i \left[ \lp 1+\frac{m_{i,R}^2+m_{j,R}^2}{2m\pi^2\beta^2}+\frac{m_im_j}{m\pi^2\beta^2}(-1)^m-\frac{m_i^2+(-1)^mm_im_j}{Q^2}\rp
\phi^{ij}_n(x)\right.\right.\nn\\
&&\left.+ (m_{i,R}^2+(-1)^mm_im_j)\frac{x}{Q^2}\phi_n'^{ij}(x)\right]\nn
\\
\nn
&&
\left.
-(-1)^{m}
		 e_jm_j \left[ \lp 1+\frac{m_{j,R}^2+m_{i,R}^2}{2m\pi^2\beta^2}+\frac{m_jm_i}{m\pi^2\beta^2}(-1)^m-\frac{m_j^2+(-1)^mm_jm_i}{Q^2}\rp
\phi^{ij}_n(1-x)\right.\right.\nn\\
\label{Whadr}
&&\left.\left.-(m_{j,R}^2+(-1)^mm_jm_i)\frac{x}{Q^2}\phi_n'^{ij}(1-x)\right]
\right\}^2\ .
\eea
The expression in the equal mass case simplifies to
\bea
&&\overline{W}
\simeq
\left(\frac{4e_im_i\pi\beta x}{Q^2}\right)^2
 \frac{1}{\left(1+\frac{M_n^2}{Q^2}x^2\right)^2}
\sum_m 
\delta\lp M_m^2-M_n^2(1-x)-Q^2\frac{1-x}{x}\rp \nn
\\
&&
\times
\left[1-(-1)^{n+m}\right]
\left\{ \lp 1-\frac{1}{m\pi^2}+\lp1+(-1)^m\rp m_i^2\lp\frac{1}{\pi^2\beta^2}-\frac{1}{Q^2}\rp\rp \phi^{ij}_n(x)+x\frac{m_{i,R}^2}{Q^2}\phi_n'^{ij}(x)\right\}^2
\ .\nn
\eea
\be
\ee
From $\overline{W}$ the component $\overline{W}^{++}$ can be obtained immediately using Eq. (\ref{W}).

At leading order in $1/Q^2$, approximating the mass of the bound state by $M_m^2 \simeq m\pi^2\beta^2$, Eq. (\ref{Whadr}) simplifies to
(note that there is a discrepancy in the relative sign of the antiparticle contribution if we compare with the 
would-be analogous expression in Ref. \cite{einhorn})
\be
\label{Wleadhad}
\overline{W}_{lead}=2\lp\frac{2x}{Q^2}\rp^2\sum_m\delta\left(m-\frac{Q^2}{\pi^2\beta^2}\frac{1-x}{x}\right)
\left[
e_im_i \phi^{ij}_n(x)-(-1)^{m}e_jm_j \phi^{ij}_n(1-x)
\right]^2\ .
\ee
At this stage one could try to approximate the sum over $m$ by an integral by 
using the EulerMacLaurin formula at the leading order, i.e. replacing $\sum_m \rightarrow \int dm$. 
This is neither mathematically justified nor can one quantify the error associated to this approximation, 
since we are dealing with Dirac deltas. For instance, it is not clear how to handle the interference term, 
which would go like $\sim \phi^{ij}_n(x)\phi^{ji}_n(x)(-1)^{\frac{Q^2(1-x)}{\pi^2\beta^2 x}}$ (though one could 
argue that it oscillates very quickly for $Q^2 \rightarrow \infty$ and away from the end points, the 
rate to which it vanishes can only be quantified when working with moments in the next section). Nevertheless, if we 
keep going and perform this naive averaging we obtain  
\be
\label{WLO}
\overline{W}^{\rm LO}=2\lp\frac{2x}{Q^2}\rp^2
\left(
\left[
e_im_i \phi^{ij}_n(x)\right]^2+\left[e_jm_j \phi^{ij}_n(1-x)
\right]^2\ 
\right),
\ee
which agrees with the expression given by Einhorn \cite{einhorn}. Note however that its analytic structure
is completely different from the one of $\overline{W}_{lead}$. On the other hand this expression will be 
useful for us in the next sections. 

\subsubsection{The forward Compton scattering amplitude}
\label{Compton}

In this section we will consider the amplitude for forward Compton scattering, which we will need for a comparison between our hadronic results and those from a calculation in perturbation theory. Let us define the tensor $T^{\mu\nu}$ in momentum space as
\bea
T^{\mu\nu}(q)&\equiv&i\int d^2xe^{iq\cdot x}
\langle ij;n|T\left\{j^{\mu}(x)j^{\nu}(0)\right\}|ij;n\rangle \nn\\
&\equiv&
\lp P_n^{\mu}-\frac{q^{\mu}q\cdot P_n}{q^2}\rp
\lp P_n^{\mu}-\frac{q^{\mu}q\cdot P_n}{q^2}\rp
 T(Q^2,x_B) 
 \,,
\eea
where in the second line we have used current conservation.
Note that due to translational invariance
\be
\label{simT}
 T^{\mu\nu}(q)= T^{\nu\mu}(-q) \ .
\ee
As the tensor structure is symmetric, this implies $T(Q^2,x_B)=T(Q^2,-x_B)$.
By using the spectral decomposition of $T^{\mu\nu}$ we obtain
\be
\text{Im} T(Q^2,x_B)= 2 \pi \widetilde{W}(Q^2,x_B)= 2\pi (\overline{W}(Q^2,x_B)+\overline{W}(Q^2,-x_B))
\,.
\ee
By using analyticity and the Cauchy theorem (and assuming that the contributions 
at infinity vanish fast enough), we can obtain the full functionality of $T(Q^2,x_B)$ from its imaginary part: let us define
\be
\nu\equiv P_n\cdot q/M_n=\frac{Q^2}{2M_n}\frac{1}{x_B} \ ,
\ee 
then we can write ($\nu_{min.}=Q^2/(2M_nx_B^{max.})$)
\bea
\label{TCauchy}
T(Q^2,x_B)&=&2\int_{\nu_{min.}}^{\infty} \frac{d\nu'^2}{\nu'^2-\nu^2-i\epsilon}\overline{W}(Q^2,\nu') \nn
\\
&=&4\int_0^{x_B^{max}}d y_B\frac{1}{y_B}\frac{\overline{W}(Q^2,y_B)}{1-\lp \frac{y_B}{x_B}\rp^2-i\epsilon}
\,,
\eea
where we have used that, since $\overline{W}(Q^2,x_B)\ne 0$ only when $x_B>0$, we only need the imaginary part 
on the positive branch of $x_B$. This means that we could obtain the expression for $T$ using only experimental data. 

\subsubsection{Moments at next-to-leading order}
\label{momDIS}

Neither $\overline{W}(Q^2, x_B)$ nor $T(Q^2,x_B)$ can be reproduced through a calculation done in perturbation theory 
for physical values of $x_B$ ($0<x_B<x_B^{max.}$). $\overline{W}$ is a sum of deltas, and its structure determines that of $T$,
but perturbation theory yields a smooth function for $\overline{W}$, as we will see shortly. A comparison between hadronic 
and perturbative results is only possible in the Deep Euclidean region, where both $Q^2$ and $x_B$ are large and a perturbative 
calculation in QCD could be justified.

$T(Q^2,x_B)$ admits an analytic expansion in $1/x_B$ for $x_B > x_B^{max.}$,
\be
\label{expxB}
 T(Q^2,x_B)=
4\sum_{N=0,2,4,...} M_N(Q^2)\frac{1}{x_B^N} \ ,
\ee
where $M_N$ are the moments of $\overline{W}$ (we take this equality also as a definition for an arbitrary $N$),
\bea
\label{MNdef}
M_N(Q^2)&\equiv& \int_0^{x_{B}^{max.}}dx_B x_B^{N-1}\overline{W}(Q^2,x_B)\nn \\
&=&\int_0^{x_{max.}} dx \frac{1+\frac{M_n^2}{Q^2}x^2}{\lp1-\frac{M_n^2}{Q^2}x^2\rp^2}\frac{x^{N-1}}{\lp1-\frac{M_n^2}{Q^2}x^2\rp^{N-1}}\overline{W}(Q^2,x)\ .
\eea
Note that only even powers of $N$ appear in $T$, due to the symmetry property shown in Eq.~ (\ref{simT}). The above expressions can 
also be rewritten as an expansion around $\nu=0$,
\be
T(\nu,Q^2)=4\sum_{N=0,2,4,...}\nu^NT^{(N)}(Q^2) \ ,
\ee
where 
\be
T^{(N)}(Q^2)\equiv \int_{\nu_{min.}}^{\infty}\frac{d\nu}{\nu^{N+1}}\overline{W}(\nu,Q^2)
=
\left(\frac{2M_n}{Q^2}\right)^NM_N(Q^2) \ .
\ee
Due to the structure of deltas in $\overline{W}$, that of the moments will be a sum over $m$. We can find an approximate result for 
this sum through the Euler-Maclaurin expansion (taking $B_2=1/6$, $B_4=-1/30\ ,\, \dots$), 
\be
\label{EulerMacLaurin}
\sum_{m=0}^{m^*}f_m=\int_0^{m^*}dm f(m) +\frac{1}{2}\lp f(0)+f(m^*)\rp +\sum_{k=1}^{\infty} \frac{B_{2k}}{(2k)!}\lp f^{(2k-1)}(m^*)-f^{(2k-1)}(0)\rp \ ,
\ee
where $f^{(n)}$ means the $n$-th derivative of the function. The limit $m=0$ corresponds to $x=x_{max.}$, and the limit $m^*$, when $m^*\to\infty$, corresponds to $x=0$. At the limit $x=0$ ($m^*\to \infty$) we can use our expression for $\overline{W}$ given in Eq. (\ref{Whadr}). As $\phi^{ij}_n(x)\stackrel{x\to 0}{\longrightarrow} c^i_n x^{\beta_i}$ both $f(\infty)$ and $f^{(n)}(\infty)$ go to zero. At the limit $x=x_{max.}$ ($m=0$) we take Eq. (\ref{1x}) to express the matrix element. With it we can see that both $f(0)$ and $f^{(n)}(0)$ are suppressed by a relative factor $O((\beta^2/Q^2)^{1+2\beta_j})$ with respect to the leading term, and so we discard them. The integral runs over all possible values of $m$, so in principle we should divide it into two regions, one in which the matrix element can be approximated by its boundary-layer expression (Eq. (\ref{-DIS})), another in which the matrix element is given by Eq. (\ref{1x}). However, no matter whether we use one expression or another, the contribution from the end-point region is suppressed again by a factor $O((\beta^2/Q^2)^{1+2\beta_j})$, due to the behavior of the wave function in that limit and the smallness of the region. So, at $O(1/Q^2)$ we just change $\sum_m \to \int dm$ in Eq. (\ref{Whadr}), and insert it inside the integral over $x$. At this order we can also use the asymptotic form of the spectrum for $M_m^2$ given in Eq. (\ref{massn2}). With these approximations the expression for the moments is
\bea
\nn
M_N^{\rm NLO}(Q^2)
&=&
\frac{8}{Q^4}
\int_0^{x_{max.}}dx\left(\frac{x}{1-\frac{M_n^2}{Q^2}x^2}\right)^N x
\left\{
e_i^2m_i^2\left(\phi_n^{ij}(x)\right)^2
+
e_j^2m_j^2\left(\phi_n^{ij}(1-x)\right)^2
\right.
\\
&&
-
2e_i^2m_i^2\phi_n^{ij}(x)\left[\frac{m_i^2}{Q^2}\phi_n^{ij}(x)-x\frac{m_{i,R}^2}{Q^2}\frac{d\phi_n^{ij}(x)}{dx}\right] \nn
\\
\nn
&&
-
2e_j^2m_j^2\phi_n^{ij}(1-x)\left[\frac{m_j^2}{Q^2}\phi_n^{ij}(1-x)-x\frac{m_{j,R}^2}{Q^2}\frac{d\phi_n^{ij}(1-x)}{dx}\right]
\\
&&
\nn
\left.
+
2e_ie_j\frac{m_i^2m_j^2}{Q^2}\left[2\frac{1-2x}{1-x}\phi_n^{ij}(x)\phi_n^{ij}(1-x)
-
x\frac{d}{dx}\left(\phi_n^{ij}(x)\phi_n^{ij}(1-x)\right)\right]
\right\} \ ,
\eea
\be
\label{MNhadr}
\ee
where the superscript NLO stands for ``next-to-leading order" and means that this expression is correct with relative $1/Q^2$ 
precision at finite $N$. We have neglected the oscillating $(-1)^m$ terms, as they give a contribution suppressed by a relative factor of $(\beta^2/Q^2)^{1+\beta_i+\beta_j}$. This is easy to see if one divides
\be
\label{evodd}
\sum_{m=0}^{\infty} (-1)^m f(m) = \sum_{m, \text{even}}f(m) -\sum_{m,\text{odd}} f(m) \ ,
\ee
and then applies the Euler-Maclaurin expansion to each separate sum. The leading contribution to the sums (the integrals) will cancel out, leaving only the subleading ones.
However, the product of two oscillating terms goes like $(-1)^{2m}=1$, which gives rise to the interference term in the last line of Eq. (\ref{MNhadr}). Being subleading in $1/Q^2$, this term was not considered in previous analysis \cite{einhorn,Batiz:2003tr}. We will see the importance of this interference term in the next section.

It must be noted that Eq. (\ref{MNhadr}) is not valid for all values of $N$. The factor $x^{N-1}$ in the definition of $M_N$ effectively selects the region of $x$ that contributes the most to the integral. This is easily seen if we express $x^N$ as
\be
x^N=e^{N \text{ln} (1- (1-x))} = e^{-N(1-x) + O((1-x)^2)} \ .
\ee
As $N\to \infty$, only the region $1-x\lesssim 1/N$ will give a sizable contribution. 
As Eq. (\ref{MNhadr}) assumes that the region $1-x\gtrsim \beta/Q$ dominates the integral, 
it is only valid for $N\lesssim Q/\beta$. To be more precise, for $N$ finite (though otherwise 
it could be large) the precision 
of our calculation is $1/Q^2$, if $N$ scales with $Q$, the precision of our computation deteriorates, 
in particular for $N \sim Q/\beta$, the precision of our computation would be $1/Q$, since 
there are (in principle) terms of ${\cal O}(N\beta^3/Q^3) \sim \beta^2/Q^2$, which we have not 
considered. 
\subsubsection{Determination of $T^{\rm NLO}$}

Since we have approximate expressions for the moments from Eq. (\ref{MNhadr}), one may think that 
(at least an approximate expression for) $T(Q^2,x_B)$ could be recovered from them using Eq. 
(\ref{expxB}):
\be
\label{TNLO}
T^{\rm NLO}(Q^2,x_B)\equiv4\sum_{N=0,2,4\dots}^{\infty} M_N^{\rm NLO}(Q^2)\frac{1}{x_B^N}
\,.
\ee
Nevertheless, this is not correct, or it rather should be quantified in which sense $T^{\rm NLO}$ provides 
with a good approximation of $T$. Since, $M_N^{NLO}$ is only valid for $N$ finite (but otherwise large), as 
compared with $Q/\beta$, Eq. (\ref{TNLO}) is only a good approximation of $T$ for $Q^2$ and $x_B$ large. This means far away from 
the physical cut ($T^{\rm NLO}$ is real in the real axis in this region). On the other hand $T^{\rm NLO}$ can be considered to be 
the generating functional for the moments with not very large $N$. Moreover, it is useful to consider $T^{\rm NLO}$ as a 
function in the $x_B$ complex plane by analytic continuation for the subsequent comparison with the computations using perturbative 
factorization. In this way $T^{\rm NLO}$ can be written in the following way
\be
\label{ImThadr}
T^{\rm NLO}(Q^2,x_B) =4\int_{0}^{x_B^{max.}}d y_B\frac{1}{y_B}\overline{W}^{\rm NLO}(Q^2,y_B)\frac{1}{1-\lp \frac{y_B}{x_B}\rp^2-i\epsilon} \ ,
\ee
where $\overline{W}^{\rm NLO}$ is given by Eq. (\ref{Whadr}), performing the substitution $\sum_m\to\int dm$ ($\overline{W}^{\rm LO}$ 
is given in Eq. (\ref{WLO})). In this equality we 
have also fixed the behavior of the function in the physical cut (the imaginary part for real $x_B$ or the $i\epsilon$ prescription) 
by demanding that it have the causality properties expected for a time-ordered propagator. Note that by 
approximating the sum over $m$ by an integral we have lost the analytic structure of the imaginary part, since $T^{\rm NLO}$
will have a continuous imaginary part, unlike that of the original $T$, which was a sum of deltas. However, as we have already mentioned, 
it is still interesting to consider the function $T^{\rm NLO}$, for in principle it should coincide with the result obtained from an 
OPE calculation, which we will perform in the next section. The function $T^{\rm NLO}$ shares with the original $T$ the trait 
that it is analytic everywhere on the complex plane except on the positive axis; therefore the resummation of the moments 
$M_N^{\rm NLO}$ amounts to the computation of the dispersion relation.

We can actually push the integration limits in Eq. (\ref{ImThadr}) to $-\infty$ and $\infty$, respectively, in terms of the 
$x$ variable. Since the 't~Hooft functions cancel out of the interval $(0,1)$, all we are doing is extending the interval 
of integration over $(x_{max},1)$. By doing so, we are introducing an error of $O(1/Q^2)^{1+2\beta_j}$, which lies anyway 
beyond the accuracy of the moments we are resumming. This way the comparison between $T^{\rm NLO}$ and $T^{OPE}$ will be clearer.

In order to give a compact, factorized expression for $T^{\rm NLO}$ we define\footnote{Note that $g_{i/j}(y)=\frac{m_{i/j}^2}{y^2}f_{i/j}(y)$, where 
$f_{i/j}(y)$ was defined in Ref. \cite{Mondejar:2008pi}. Note as well that the definitions of $J_i$ are, accordingly, also 
slightly different here and in Ref. \cite{Mondejar:2008pi}. The difference between both definitions has to do on whether 
one chooses the ``+" or ``-" component for the distribution amplitudes. In particular, $f_{i/j}$ could be obtained from 
$g_{i/j}$ applying twice the equations of motion to their Fourier transform with the proper normalization. 
On the other hand the definition of $g_{int.}=f_{int.}$ and $J_{int.}$ are equal.} the following functions 
(sums over the color indices of the fields are implicit):

\bea
\label{fi}
g_i(y)&\equiv& \frac{m_i^2}{y^2}\phi_n^2(y)=(P_n^+)^2\int\frac{dx^-}{2(2\pi)}e^{-iy\frac{P_n^+x^-}{2}}\\
&&\times\lp\langle n| \psi_{i,-}^{\dagger}(x^-)\Phi(x^-,0)\psi_{i,-}(0)|n\rangle+\frac{\beta^2}{m_i^2}{}^{(0)}
\langle n| \psi_{i,-}^{\dagger}(0)\psi_{i,-}(0)|n\rangle^{(0)} \rp\nn\\
\label{fj}
g_j(y)&\equiv&\frac{m_j^2}{y^2}\phi_n^2(1-y)=-(P_n^+)^2\int\frac{dx^-}{2(2\pi)}e^{-iy\frac{P_n^+x^-}{2}}\\
&&\times\lp \langle n| \psi_{j,-}^{\dagger}(0)\Phi(0,x^-)\psi_{j,-}(x^-)|n\rangle+\frac{\beta^2}{m_j^2}{}^{(0)}
\langle n| \psi_{j,-}^{\dagger}(0)\psi_{j,-}(0)|n\rangle^{(0)}\rp \nn
\eea
\bea
 \label{fint}
&&g_{int.}(y)\equiv\frac{m_im_j}{y(1-y)}\phi_n^{ij}(y)\phi_n^{ij}(1-y)=\frac{(P_n^+)^2}{N_c}
\int_{-\infty}^{\infty} \frac{dx^-}{2(2\pi)}e^{-iy P_n^+ \frac{x^-}{2}} \\
&&\qquad\quad\times\int_{-\infty}^{\infty} dz^- \langle ij;n|\psi^{\dagger}_{i,-}(x^-)\Phi(x^-,z^-)
\psi_{j,-}(z^-)\psi^{\dagger}_{j,+}(0)\Phi(0,z^-)\psi_{i,+}(z^-)|ij;n\rangle \ ,\nn
\eea
where as before the Wilson lines, trivial in the light-cone gauge, have been inserted to make gauge invariance explicit. 
The expressions for $g_i$ can be compared with those shown in Ref. \cite{Burkardt:1991hu}. We find agreement with them. 

The $\beta^2/m_{i}^2$ terms in Eqs. (\ref{fi}) and (\ref{fj}) have been inserted to cancel the ``off-diagonal" contribution 
to the matrix elements $\langle n| \psi_{i,-}^{\dagger}(x^-)\Phi(x^-,0)\psi_{i,-}(0)|n\rangle$ and \newline
$\langle n| \psi_{j,-}^{\dagger}(0)\Phi(0,x^-)\psi_{j,-}(x^-)|n\rangle$. Note however that this ``off-diagonal" contribution does not 
correspond to the matrix element in Eq. (\ref{offdiag}). The term $b_fa_f$ in the current, from which the matrix element 
in Eq. (\ref{offdiag}) comes, gives a null contribution in this case. This is due to the insertion of an external momentum 
$P_n^+y$. The ``off-diagonal" contribution originates in this case from the term $a_f^{\dagger}b_f^{\dagger}$.
The matrix element in Eq. (\ref{fint}) does not have ``off-diagonal" contributions: the ``off-diagonal" contribution involves 
a two-meson intermediate state, as in Fig. \ref{figmatelem}, which is incompatible with the color structure of the matrix element.

The functions $g(y)$ encode non-perturbative information. We can write $T^{\rm NLO}$ in terms of these functions times some other functions $J$ which will hold the perturbative contribution,
\be
\label{factTm}
T^{\rm NLO}(Q^2,x_B) = -2\lp\frac{4}{Q^2}\rp^2\int_{-\infty}^{\infty} dy \left\{ e_i^2J_i(x,y) g_i(y)+e_j^2J_j(x,y) g_j(y)+ e_ie_j J_{int.}(x,y)g_{int.}(y)\right\} \ ,
\ee
where the functions $J$ are defined as
\bea
\label{JX}
J_i(x,y)&\equiv&\left[x^2\lp 1-2\frac{m_i^2}{Q^2}-2\frac{M_n^2}{Q^2}y^2\rp+x^3\frac{m_{i,R}^2}{Q^2}\frac{d}{dx}\right]\frac{y^3}{y^2-x^2+i\epsilon} \nn\\
J_j(x,y)&\equiv&\left[x^2\lp 1-2\frac{m_j^2}{Q^2}-2\frac{M_n^2}{Q^2}y^2\rp+x^3\frac{m_{j,R}^2}{Q^2}\frac{d}{dx}\right]\frac{y^3}{y^2-x^2+i\epsilon} \nn\\
J_{int.}(x,y)&\equiv& 2\frac{m_im_j}{Q^2}\left[2x^2(1-2y)-x^3(1-y)\frac{d}{dx}\right]\frac{y^2}{y^2-x^2+i\epsilon} \ .
\eea
This is the factorized form we expect from an OPE. However, as we will see shortly, the interference contribution (the term involving $g_{int.}$) lies beyond the domain of the OPE.

The functions $f$ are real, so the expression for $\text{Im}T^{\rm NLO}$ has also a factorized form,
\bea
\text{Im}T^{\rm NLO}(Q^2,x_B) &=& -2\lp\frac{4}{Q^2}\rp^2\int_{-\infty}^{\infty} dy \left\{ e_i^2\text{Im}[J_i(x,y)] g_i(y)+e_j^2\text{Im}[J_j(x,y)] g_j(y)\right.\nn \\
&&\qquad\qquad\qquad\qquad\left.+ e_ie_j \text{Im}[J_{int.}(x,y)]g_{int.}(y)\right\} \ .\nn
\eea
\be
\ee

Recall that we obtained the moments $M_N^{\rm NLO}$, and therefore $T^{\rm NLO}$, through the use of the 
dispersion relation given in Eq. (\ref{TCauchy}), relating the discontinuity of $T(Q^2,\nu)$ on the positive 
axis to its structure anywhere else on the complex plane. Therefore, the functions $J$ are actually defined as 
(changing variables from $J(x,y)$ to $J(x,\nu)$)
\be
\label{defJ}
J(x,\nu)\equiv \int_{0}^{\infty} d\nu'^2\frac{\frac{1}{\pi}\text{Im} J(x,\nu')}{\nu'^2-\nu^2-i\epsilon} \ .
\ee
There is non-trivial information in this equation: the result from a direct calculation 
of $J(x,\nu)$ in the Euclidean and the result from the dispersion relation might differ in a polynomial 
(see e.g. \cite{deRafael:1998}). The reason we use this prescription is that 
we are interested in the comparison between the hadronic and OPE results, and we will use 
the same prescription in the perturbative computation. In this way we aim to eliminate spurious differences 
between both computations due to dispersion-relation issues.

\subsubsection*{Expression of $M_N^{\rm NLO}$ in terms of matrix elements}

The expression for the moments given in Eq. (\ref{MNhadr}) can be rewritten in terms of 
matrix elements, expectation values of some operators. We expect in this way to rewrite the 
moments in terms of an OPE expansion. However, not all the matrix elements that will 
appear can correspond to an OPE expansion: the interference term can only be represented 
through non-local matrix elements, 

For simplicity's sake we will keep the factors of $M_n^2/Q^2$ explicit. They are actually not 
relevant for our purpose, which is spotting differences between full and OPE results.
Although there is some dynamical, non-perturbative information encoded in $M_n^2$, the presence of 
these factors in Eq. (\ref{MNhadr}) has a kinematical origin: they come from our definition of 
$\overline{W}$ (Eq. (\ref{defW})) and the change of variables from $x_B$ to $x$ (Eq. (\ref{xeqxB})). 
Therefore, they will also be present in our later OPE calculation in exactly the same way. 

Starting either from Eq. (\ref{MNhadr}) or from Eq. (\ref{factTm}), using the definitions given 
in Eqs. (\ref{fi})-(\ref{fint}), and integrating by parts we can write $M_N^{\rm NLO}$ in terms of 
matrix elements with derivatives inserted between the fields. The expression at $O(1/Q^2)$ is
\bea
&&M_N^{\rm NLO}(Q^2)=\frac{8}{Q^4}\left\{e_i^2
\langle ij;n| \psi_{i,-}^{\dagger}(0)\frac{(i\overleftrightarrow{D}^+)^{N+3}}{(P_n^+)^{N+2}}\right.\nn\\
&&\qquad\qquad\qquad\quad\quad\times\lp 1 -2\frac{m_i^2}{Q^2} - (N+2)\frac{m_{i,R}^2}{Q^2}+N\frac{M_n^2}{Q^2}\frac{(i\overleftrightarrow{D}^+)^2}{(P_n^+)^2}\rp \psi_{i,-}(0)| ij;n\rangle \nn\\
&&\qquad\qquad\qquad\quad-e_j^2
\langle ij;n| \psi_{j,-}^{\dagger}(0)\frac{(-i\overleftrightarrow{D}^+)^{N+3}}{(P_n^+)^{N+2}}\nn\\
&&\qquad\qquad\qquad\quad\quad\times\lp 1 -2\frac{m_j^2}{Q^2} - (N+2)\frac{m_{j,R}^2}{Q^2}+N\frac{M_n^2}{Q^2}\frac{(-i\overleftrightarrow{D}^+)^2}{(P_n^+)^2}\rp \psi_{j,-}(0)| ij;n\rangle\nn \\
&&+\frac{2e_ie_j}{N_c}\frac{m_im_j}{Q^2}\int dz^- \langle ij;n| \psi_{i,-}^{\dagger}(0) \frac{(-i\overleftarrow{D}^+)^{N+2}}{(P_n^+)^{N+1}}\left[(N+4)\lp 1-\frac{-i\overleftarrow{D}^+}{P_n^+}\rp-2 \frac{-i\overleftarrow{D}^+}{P_n^+}\right] \nn\\
&&\left.\times\Phi(0,z^-)\psi_{j,-}(z^-)\psi_{j,+}^{\dagger}(0)\Phi(0,z^-)\psi_{i,+}(z^-)|ij;n\rangle 
\right\} \ , \nn
\eea
\be
\label{MNmat}
\ee
where $\psi^{\dagger}(x)(\overleftarrow{D}^+) \equiv \lp D^+ \psi(x)\rp^{\dagger}$, 
and $\overleftrightarrow{D} \equiv \frac{1}{2} (\overrightarrow{D}-\overleftarrow{D})$. 

The interference term is represented by the matrix element in the last two lines. 
The origin of the interference term is the constructive interference between two oscillating terms, 
$(-1)^m\cdot (-1)^m=(-1)^{2m}=1$. These oscillating terms give, after summing over $m$, an 
$O(\beta^2/Q^2)^{1+\beta_i+\beta_j}$ contribution to the moments, but their interference is enhanced.
A non-analytic dependence on $1/Q^2$ like that of the oscillating terms seems out of the reach of an OPE. Note that 
for large $m$, one may think of 
\be
(-1)^m \rightarrow e^{i\frac{Q^2}{\pi\beta^2}\frac{1-x}{x}}
\ee
which has a non-analytic expansion in $1/\beta$. 
So, although the interference term is formally a simple NLO term in an $1/Q^2$ expansion, it is built out of terms 
which seem to be beyond an OPE expansion, whose non-OPE nature would survive in the form of a non-local matrix element. 
Either way, a complete understanding of this non-local 4-field correlator 
is still lacking, but what is certain is that because of its non-local nature it is beyond an OPE expansion. 

The problem we are encountering might be enhanced by the large $N_c$ limit: the $(-1)^m$ terms arise from the null 
width of the resonances. Incorporating finite widths to the resonances (going to higher 
orders in the $1/N_c$ expansion), oscillations would be milder, perhaps moving this interference down to some higher order of $1/Q^2$. 
In any case, our result seems to indicate a breakdown of the OPE for DIS in the 't~Hooft model.

\subsection{Perturbative factorization}

In this section we will compute the amplitude for the forward Compton scattering 
following the recipe of perturbative factorization. We will first compute the imaginary 
part of $T^{--}$, from which we will find the imaginary part of $T^{OPE}$, and then the full $T^{OPE}$ 
through a dispersion relation. Perturbation theory only makes sense in the Deep Euclidean domain, 
but if $T^{OPE}$ is an analytic function, we can obtain its behavior in that region from its discontinuity 
in the positive axis, irrespective of whether this function properly describes $T$ in the physical cut or not.

We will compare our result for $T^{OPE}$ with Eq. (\ref{factTm}). 
From $T^{OPE}$ we will derive the coefficients $M_N^{OPE}$, which we will check against Eq. (\ref{MNmat}).

\subsubsection{Calculation of $T^{OPE}$}
\label{pertDIS}
$T^{--}$ is defined as
\be
T^{--}= i\sum_{h,h'}e_he_{h'}\int d^2xe^{iq\cdot x}
\langle ij,n|T\left\{\bar{\psi}_h(x)\gamma^-\psi_h(x) \bar{\psi}_{h'}(0)\gamma^-\psi_{h'}(0)\right\}|ij;n\rangle \ .
\ee
In perturbation theory, at leading order in $\beta^2$, $T^{--}$ reads

\bea
\label{partT}
T^{--}_{OPE}&=&4i \int d^2x e^{iq\cdot x}
\left\{e_i^2\left[ P_i^+(x) \langle ij;n|\psi^{\dagger}_{i,-}(x)\Phi(x,0)\psi_{i,-}(0)|ij;n\rangle\right.\right. \nn\\
&&\left.\quad\qquad\qquad\qquad+ P_i^+(-x)  \langle ij;n|\psi^{\dagger}_{i,-}(0)\Phi(0,x)\psi_{i,-}(x)|ij;n\rangle\right]\nn\\
&&\quad\qquad\qquad\quad+e_j^2\left[ P_j^+(-x)  \langle ij;n|\psi^{\dagger}_{j,-}(0)\Phi(0,x)\psi_{j,-}(x)|ij;n\rangle\right.\nn\\
&&\quad\qquad\qquad\qquad\,\left.\left.+  P_j^+(x)  \langle ij;n|\psi^{\dagger}_{j,-}(x)\Phi(x,0)\psi_{j,-}(0)|ij;n\rangle\right]\right\}\ ,
\eea

where $P^+_{i}(x)$ is the ``+" component of the free quark propagator,
\be
\label{Pplus}
P^+_{i}(x)\equiv \frac{1}{2}Tr[\gamma^-P_{i}^f(x)]=
\int \frac{d^2k}{(2\pi)^2}e^{-ikx}\frac{m_i^2}{k^2}\frac{i}{k^+-\frac{m_{i}^2}{k^-} +i\frac{\epsilon}{k^-}} \ .
\ee
We have written the Wilson lines to restore gauge invariance. 
These Wilson lines include both components of the gluon field, $A^+$ and $A^-$, in an obvious generalization 
of their definition in Eq. (\ref{Wline}).

\begin{figure}[hpt]
\begin{center}
\includegraphics[width=0.9\columnwidth]{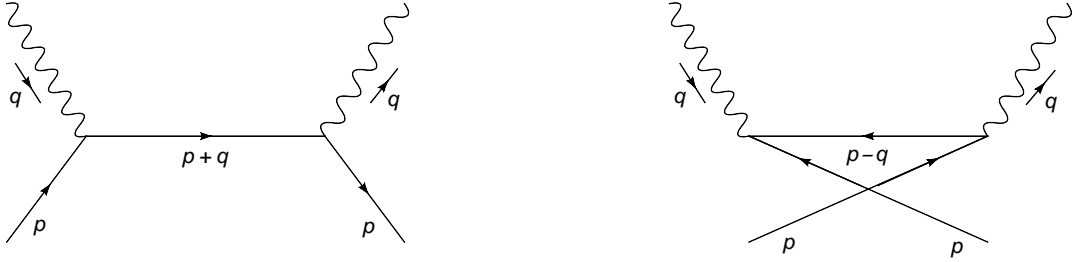}
\caption{\it The direct and crossed diagrams representing, in momentum space, the two first terms in 
Eq. (\ref{partT}). The momentum $p$ is the momentum of the quark inside the meson $n$. } 
\label{parton}
\end{center}
\end{figure}

The picture for the process represented by Eq. (\ref{partT}) is shown in Figure \ref{parton}, for the particle case 
(the antiparticle case just involves switching around all the arrows in the quark lines).

In order to deal with the matrix elements we take advantage of the kinematics we have chosen: in our frame, 
$q^-\to \infty$. Whether we consider the direct or the crossed diagram, this implies that the quark propagating 
between the two vertices has a very large $p^-$ component, that is to say, its propagation takes place in a very 
short ``time" ($x^+$ takes the role of time in our quantization frame). We can therefore expand the matrix elements in powers of $x^+$,
\bea
\label{exp}
\langle ij;n|\psi^{\dagger}_{i,-}(x)\Phi(x,0)\psi_{i,-}(0)|ij;n\rangle&\simeq& \langle ij;n|\psi^{\dagger}_{i,-}(x^-,0)\Phi(x^-,0)\psi_{i,-}(0)|ij;n\rangle \\
&&+ \frac{1}{2} \langle ij;n|\psi^{\dagger}_{i,-}(x^-,0)\overleftarrow{D}^-\Phi(x^-,0)\psi_{i,-}(0)|ij;n\rangle x^+  \ .\nn
\eea
Effectively, by letting the coordinate $x^-$ untouched and expanding in $x^+$, what we are doing is resuming all powers of $p^+$ 
in an OPE expansion of the amplitude (recall that $\frac{\partial}{\partial x^-}\sim p^+$). Since we stay at $O(1/Q^2)$ we don't 
have to worry about the term in the second line in Eq. (\ref{exp}): due to the $x^+$ multiplying at the right its contribution will 
be suppressed by a relative factor $1/Q^4$. 

Using the definitions of Eqs. (\ref{fi}) and (\ref{fj}) we can reexpress the leading matrix elements as
\be
\label{fi2}
\langle ij;n|\psi^{\dagger}_{i,-}(x^-,0)\Phi(x^-,0)\psi_{i,-}(0)|ij;n\rangle=\frac{1}{P_n^+}\int dy 
e^{i y \frac{P_n^+x^-}{2}}g_i(y)-\frac{\beta^2}{m_i^2}{}^{(0)}\langle n| \psi_{i,-}^{\dagger}(0)\psi_{i,-}(0)|n\rangle^{(0)}
\ee
\be
\label{fj2}
\langle ij;n|\psi^{\dagger}_{j,-}(0)\Phi(0,x^-)\psi_{j,-}(0,x^-) |ij;n\rangle
= -\frac{1}{P_n^+}\int dy e^{i y \frac{P_n^+x^-}{2}}g_j(y)
-\frac{\beta^2}{m_j^2}{}^{(0)}\langle n| \psi_{j,-}^{\dagger}(0)\psi_{j,-}(0)|n\rangle^{(0)} \ .
\ee

In writing $\text{Im}{T}^{--}_{OPE}$, one could neglect the $\beta^2/m_{i}^2$ terms. 
The reason is that its Fourier transform is proportional to $\delta(y)$ and, 
after we sum the direct and crossed contributions, it will get multiplied by powers of $y$, 
giving a vanishing contribution to $\text{Im}{T}^{--}_{OPE}$. Thus, effectively, we are 
representing the matrix elements through some parton distribution functions $g_{i,j}(y)$. We write the imaginary part of $T^{--}_{OPE}$ as 
\bea
\label{ImTpertlead}
\text{Im}{T}^{--}_{OPE}&=&-\frac{4e_i^2}{(P_n^+)^2}\int_{-\infty}^{\infty} dy   g_i(y) \,\text{Im}\lp \frac{1}{y-x\lp 1+
\frac{m_i^2}{Q^2}-i\epsilon\rp}+  \frac{1}{y+x\lp1+\frac{m_i^2}{Q^2}-i\epsilon\rp} \rp\nn\\
&&-\frac{4e_j^2}{(P_n^+)^2}\int_{-\infty}^{\infty} dy  g_j(y)\,\text{Im}\lp \frac{1}{y-x\lp 1+
\frac{m_j^2}{Q^2}-i\epsilon\rp}+  \frac{1}{y+x\lp1+\frac{m_j^2}{Q^2}-i\epsilon\rp} \rp\nn\\
&=&-\frac{8e_i^2}{(P_n^+)^2}\int_{-\infty}^{\infty} dy  g_i(y)\,\text{Im}\lp\frac{y}{y^2-x^2\lp 1+2\frac{m_i^2}{Q^2}\rp +i\epsilon}\rp\nn\\
&&- \frac{8e_j^2}{(P_n^+)^2}\int_{-\infty}^{\infty} dy  g_j(y)\,\text{Im}\lp\frac{y}{y^2-x^2\lp 1+2\frac{m_j^2}{Q^2}\rp +i\epsilon}\rp\ .
\eea

\begin{figure}[hhh]
\begin{center}
\includegraphics[width=0.5\columnwidth]{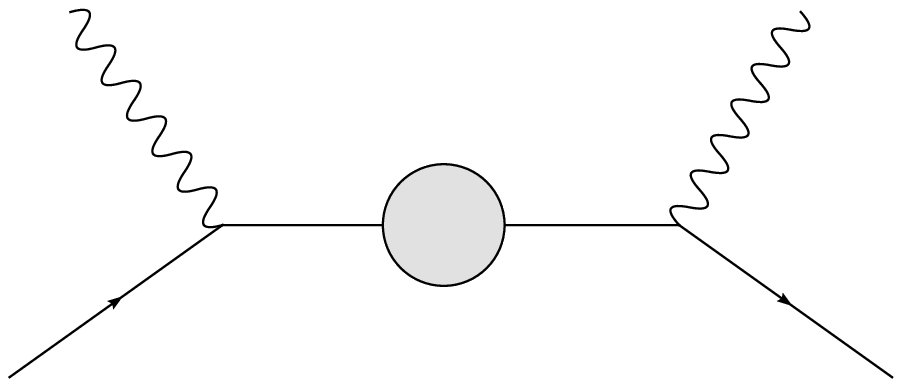}
\end{center}
\makebox[1.0cm]{\phantom b}
\put(-30,10){\epsfxsize=8truecm \epsfbox{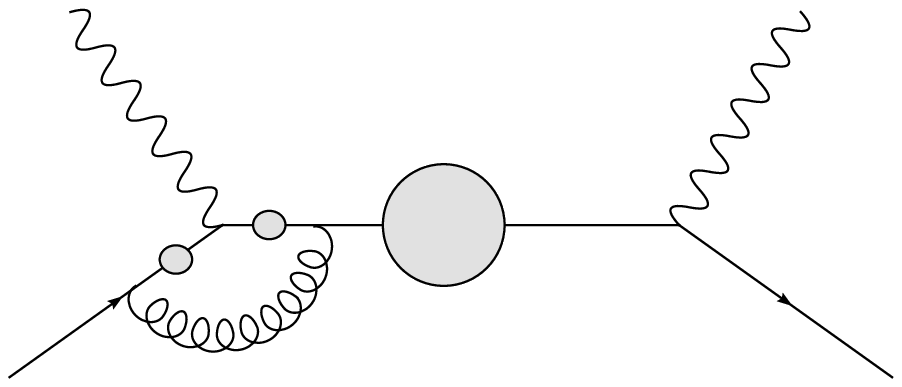}}
\put(210,10){\epsfxsize=8truecm \epsfbox{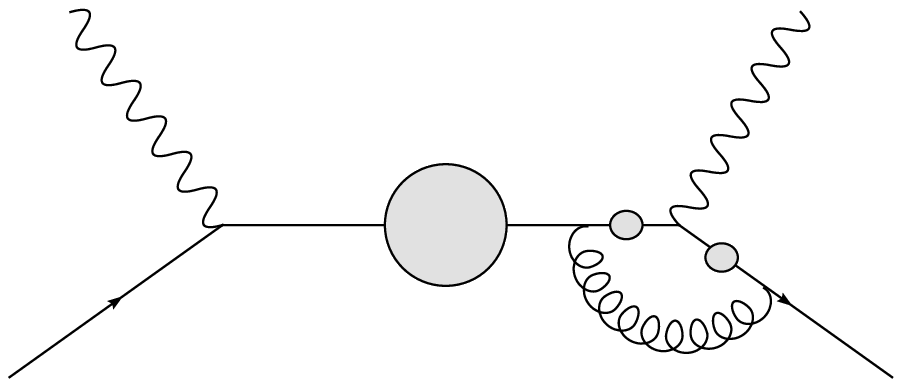}}
\vspace{1 mm}
\caption{\it Diagrams contributing to the perturbative compuation at $O(\beta^2/Q^2)$. }
\label{figb}
\end{figure}

\begin{figure}[hhh]
\begin{center}
\includegraphics[width=0.9\columnwidth]{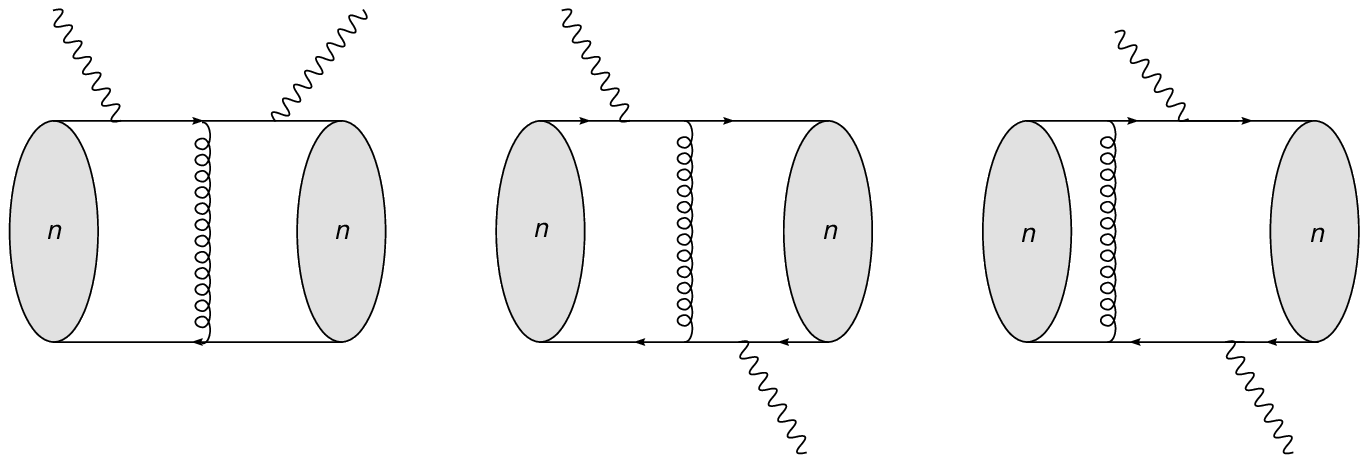}
\caption{\it Additional $\beta^2$ corrections.}
\label{agb}
\end{center}
\end{figure}

Next we consider corrections in $\beta^2$. Actually, we will write these corrections together with the leading order result we have 
just obtained in a combined single expression. We want to compute the following matrix element at $O(\beta^2/Q^2)$:
\be
\int d^2xe^{iq \cdot x}\langle p| T\{\bar{\psi}(x)\gamma^-\psi(x)\bar{\psi}(0)\gamma^-\psi(0)\}|p\rangle \ ,
\ee
where $|p\rangle\equiv a^{\dagger}(p)|0 \rangle$ (for the moment we do not specify the value of $p^2$). 
We have to consider the diagrams in Figure \ref{figb}, 
where the ``blobs" represent the renormalized propagators to all orders in $\beta^2$. We only show the 
direct contribution, the crossed-diagram contributions can be obtained from it with the change $x \rightarrow -x$.
A similar computation should also be carried out for the antiparticle contribution. Other possible corrections are 
given by the diagrams in Figure \ref{agb} (plus the symmetric ones), but these are suppressed by a relative order 
of $1/Q^4$.

The first diagram gives
\bea
\parbox{30mm}{
\includegraphics[width=0.25\columnwidth]{blob.eps}
}\qquad\quad
&=&4 \frac{m_i^2}{(p^++q^+)(p^-+q^-)}\frac{m_i^2}{(p^+)^2}\frac{1}{p^++q^+-\frac{m_{i,R}^2}{p^-+q^-}+i\epsilon} \nn \\
&\to&4 \lp 1+\frac{\beta^2}{m_i^2}\rp\frac{m_i^2}{(p^+)^2}\frac{1}{p^++q^+-\frac{m_{i,R}^2}{p^-+q^-}+i\epsilon}
\ ,
\eea
where in the last line we have applied the momentum conservation delta to the factor 
$m_i^2/((p^++q^+)(p^-+q^-))$ (to order $\beta^2$), since we are only interested in the imaginary part of this diagram.

The vertex correction reads
\be
\label{vb1}
\parbox{30mm}{
\includegraphics[width=0.2\columnwidth]{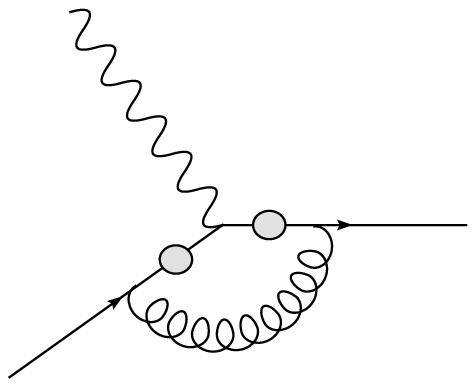}
}\quad
=\gamma^+\frac{m_i^2}{(p^+)^2}\frac{\beta^2}{Q^2} f\lp \frac{-q^+}{p^+}\rp \ ,
\ee
where we have defined
\bea
\label{fint1l}
f\lp  z\rp
&=&
z\int_{1-z}^1\frac{dy}{y^2}\frac{1}{(1-y)(1-z-y)-\frac{m_{i,R}^2}{Q^2}z^2}
\\
\nn
&=&
\left[-2\lp 2- z\lp 2 - z- 2\frac{m_i^2}{Q^2} z\rp\rp \text{Arctanh}\lp \frac{1}{\sqrt{1+4\frac{m_{i,R}^2}{Q^2}}}\rp \right.\nn \\
&&\left.+ \sqrt{1+4\frac{m_{i,R}^2}{Q^2}}z\lp  z \frac{1- z-\frac{m_i^2}{Q^2}\lp z\rp^2}{1- z} - \lp2- z\rp\text{ln}\lp1- z\rp\rp\right]\nn\\
&&\times\frac{1}{\sqrt{1+4\frac{m_{i,R}^2}{Q^2}}\lp 1-z-\frac{m_{i,R}^2}{Q^2}\lp z\rp^2\rp^2} \nn \ .
\eea
\be
\ee
We then find
\bea
\parbox{30mm}{
\includegraphics[width=0.25\columnwidth]{vblob1.eps}
}\qquad\quad
&=&4\frac{m_i^2}{(p^++q^+)(p^-+q^-)}\frac{m_i^2}{(p^+)^2}\frac{p^++q^+}{p^+}\frac{\beta^2}{Q^2}
\label{vb2}
\\
&&
\times
f\lp  
\frac{-q^+}{p^+}\rp 
\frac{1}{p^++q^+-\frac{m_{i,R}^2}{p^-+q^-}+i\epsilon} \nn\\
&\to&4\frac{1}{2}\lp -\frac{\beta^2}{m_i^2} -2\frac{\beta^2}{Q^2}\rp \frac{m_i^2}{(p^+)^2}\frac{1}{p^++q^+-\frac{m_R^2}{p^-+q^-}+i\epsilon} \ , \nn
\eea
where in the last line we have applied the momentum conservation delta coming from the intermediate propagator and expanded at 
order $\beta^2$ (this is the relevant result if we only want the imaginary part). Within this approximation we have 
$m^2/((p^++q^+)(p^-+q^-)) \simeq 1$ and 
\be
(1+q^+/p^+)\beta^2/Q^2f(-q^+/p^+) \simeq - \frac{\beta^2}{2m_{i,R}^2}\left(1+\frac{5}{3}\frac{m_{i,R}^2}{Q^2}+\frac{1}{3}\frac{p^2}{Q^2}\right)
\simeq - \frac{\beta^2}{2m_{i}^2}\left(1+2\frac{m_{i}^2}{Q^2}\right)
\ee
where in the last line we have only kept terms of order $\beta^2$ and approximated $p^2 \simeq m^2$. It is interesting to discuss where 
this contribution comes from in the original integral in Eq. (\ref{fint1l}). Due to the delta of conservation $z \sim 1$ and one can rewrite 
$z \sim 1-\delta$. Then the integral has contributions from $y \sim 1$ and $y \sim \delta \ll 1$. With the precision of our computation only the 
region $y \sim \delta$ contributes to the integral (both regions would start to mix at order $1/Q^4$). 

We can finally write the perturbative result for Im$T^{--}$. It reads
\bea
\text{Im}T^{--}_{OPE}&=&-\frac{8e_i^2}{(P_n^+)^2}\int_{-\infty}^{\infty} dy  g_i(y)\,
\text{Im}\left[ y\lp 1-2\frac{\beta^2}{Q^2}\rp\frac{1}{y^2-x^2\lp 1+\frac{m_{i_R}^2}{Q^2}\rp^2 +i\epsilon}\right]\\
&&- \frac{8e_j^2}{(P_n^+)^2}\int_{-\infty}^{\infty} dy  g_j(y)\,
\text{Im}\left[ y\lp 1-2\frac{\beta^2}{Q^2}\rp\frac{1}{y^2-x^2\lp 1+\frac{m_{j,R}^2}{Q^2}\rp^2 +i\epsilon}\right]\ , \nn
\eea
or (strictly at $O(\beta^2/Q^2)$)
\bea
\text{Im}T^{--}_{OPE}&=&-\frac{8e_i^2}{(P_n^+)^2}\int_{-\infty}^{\infty} dy  g_i(y)\,\text{Im}
\left[\lp y\lp 1-2\frac{\beta^2}{Q^2}\rp-2\frac{\beta^2}{Q^2}x^2\frac{d}{dx^2}\rp\frac{1}{y^2-x^2\lp 1+\frac{m_{i}^2}{Q^2}\rp^2 +i\epsilon}\right]\nn\\
&&- \frac{8e_j^2}{(P_n^+)^2}\int_{-\infty}^{\infty} dy  g_j(y)\,\text{Im}
\left[\lp y\lp 1-2\frac{\beta^2}{Q^2}\rp-2\frac{\beta^2}{Q^2}x^2\frac{d}{dx^2}\rp\frac{1}{y^2-x^2\lp 1+\frac{m_{j}^2}{Q^2}\rp^2 +i\epsilon}\right]\ . \nn
\eea
\be
\label{T--pert}
\ee

From $\text{Im}T_{OPE}^{--}(Q^2,x)$ we can obtain $\overline{W}_{OPE}(Q^2,x)$, 
\bea
\label{WOPE}
&&
\overline{W}_{OPE}(Q^2,x)=\lp\frac{2x}{q^-}\rp^2\frac{1}{\lp 1+\frac{M_n^2}{Q^2}x^2\rp^2}\frac{1}{2\pi}\text{Im}T^{--}_{OPE}|_{x>0}\\
&&=\frac{8}{Q^4}x^4\frac{1}{\lp 1+\frac{M_n^2}{Q^2}x^2\rp^2} \lp 1 -2\frac{\beta^2}{Q^2}\rp\left[ e_i^2 g_i\lp x\lp 1+\frac{m_{i,R}^2}{Q^2}\rp\rp+e_j^2 g_j\lp x\lp 1+\frac{m_{j,R}^2}{Q^2}\rp\rp\right] \ .\nn
\eea
Now, using dispersion relations (the analogous to Eq. (\ref{TCauchy}) but with $\pm \infty$ integration limits, since 
we do not need to know for which values of $x$ the integrand gives a non-zero contribution, this is built-in in the result), 
expanding and integrating by parts, our result for $T^{OPE}$ at $O(1/Q^2)$ is
\be
\label{TOPE}
T^{OPE}(Q^2,x_B) = -2\lp\frac{4}{Q^2}\rp^2\int_{-\infty}^{\infty} dy \left\{ e_i^2J_i(x,y) g_i(y)+e_j^2J_j(x,y) g_j(y)\right\} \ .
\ee

Eq. (\ref{TOPE}) can be compared with our factorized expression for $T^{\rm NLO}$ given in Eq. (\ref{factTm}).
We recognize the particle and antiparticle contributions, but the interference term is missing.

The corrections involving the exchange of a gluon between particle and antiparticle represented in 
Figure \ref{agb} (plus the symmetric ones) are suppressed by a factor of $1/Q^4$. Perturbation theory does 
indeed contemplate interference terms, but they cannot account for $g_{int.}$.

\begin{figure}[hhh]
\begin{center}
\includegraphics[width=0.57\columnwidth]{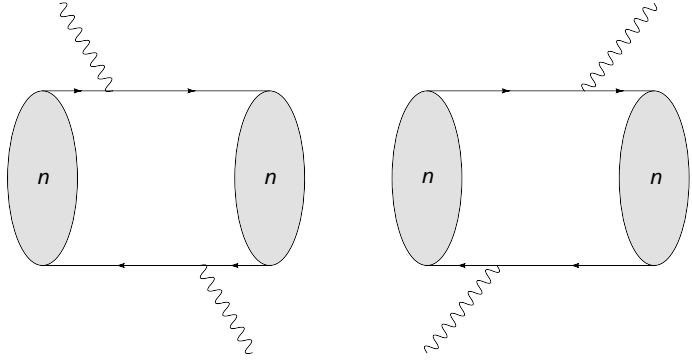}
\caption{\it Additional (non-perturbative) diagrams involving both the particle and the antiparticle.}
\label{ab}
\end{center}
\end{figure}

Besides these perturbative corrections there are also the diagrams shown in Figure \ref{ab}, which are suppressed by a relative 
factor of $(\beta^2/Q^2)^{1+\beta_i}$ and $(\beta^2/Q^2)^{1+\beta_j}$, respectively. Strictly speaking they do not belong to 
the dominion of perturbation theory, but they illustrate the fact that the interference term cannot be produced by any diagrammatic 
calculation at finite order in $\beta^2$, reinforcing the idea that our interference term, although formally an $1/Q^2$ term, is non-OPE in nature.

From our OPE calculation we see that we can understand the perturbative functions $J_{i}$ and $J_j$ as coming from the 
propagator of a collinear quark times kinematical and loop corrections (after an expansion). 
The connection between the functions $J$ and the quark propagator that appears in Eq. (\ref{TOPE}) is however not direct 
in general. Note, in particular, that the quark propagator may depend on the quantization frame and the gauge fixing used, which 
makes the explicit expression of the quark propagator different. We have tried to avoid as much as possible these ambiguities by using dispersion 
relations and demanding the $T^{\mu\nu}$ to have the expected tensor structure. 

\subsubsection{Moments}
\label{mompertDIS}

From Eq. (\ref{TOPE}) we can find the expression for the OPE coefficients $M_N^{OPE}$. Actually, for this purpose it is more convenient to keep the factors of $\frac{m_{x,R}^2}{Q^2}$ inside the functions $f_{i,j}$ as in Eq. (\ref{WOPE}).
The expression for the moments obtained from a partonic approach is then
\bea
\label{MNpart}
M_N^{OPE}(Q^2)
&=&
\frac{8}{Q^4}
\int_{-\infty}^{\infty}dx\left(\frac{x}{1-\frac{M_n^2}{Q^2}x^2}\right)^Nx^3\lp 1-2\frac{\beta^2}{Q^2}\rp \nn\\
&&\times\left[ e_i^2 g_i\lp x\lp 1+\frac{m_{i,R}^2}{Q^2}\rp\rp+e_j^2 g_j\lp x\lp 1+\frac{m_{j,R}^2}{Q^2}\rp\rp\right] \ ,\nn\\
\eea
which, using the definitions of $g_{i,j}$ given in Eqs. (\ref{fi}) and (\ref{fj}) and integrating by parts, can be reexpressed as
\bea
\label{MNmatpert}
&&M_N^{OPE}(Q^2)=\frac{8}{Q^4}\left\{e_i^2 
\langle ij;n| \psi_{i,-}^{\dagger}(0)\frac{(i\overleftrightarrow{D}^+)^{N+3}}{(P_n^+)^{N+2}}\right.\nn\\
&&\qquad\qquad\qquad\quad\quad\times\lp 1 -2\frac{m_i^2}{Q^2} - (N+2)\frac{m_{i,R}^2}{Q^2}+N\frac{M_n^2}{Q^2}\frac{(i\overleftrightarrow{D}^+)^2}{(P_n^+)^2}\rp \psi_{i,-}(0)| ij;n\rangle \nn\\
&&\qquad\qquad\qquad\quad-e_j^2
\langle ij;n| \psi_{j,-}^{\dagger}(0)\frac{(-i\overleftrightarrow{D}^+)^{N+3}}{(P_n^+)^{N+2}}\nn\\
&&\left.\qquad\qquad\qquad\quad\quad\times\lp 1 -2\frac{m_j^2}{Q^2} - (N+2)\frac{m_{j,R}^2}{Q^2}+N\frac{M_n^2}{Q^2}\frac{(-i\overleftrightarrow{D}^+)^2}{(P_n^+)^2}\rp \psi_{j,-}(0)| ij;n\rangle\right\} \ .\nn\\
\eea
If we compare $M_N^{\rm NLO}$ with $M_N^{OPE}$ we confirm that the OPE does indeed get the terms involving local matrix elements right, 
but cannot grasp the non-local one.

\section{Conclusions}
\label{concDIS}

We have thoroughly studied the 't Hooft model. We have obtained exact 
expressions for the current matrix elements in terms of the 't Hooft wave function. 
We have then studied the Deep Inelastic Scattering of a lepton off a meson in the 't~Hooft model. 
We have calculated the full, non-perturbative expression of $\overline{W}^{\mu\nu}$, and 
observed maximal duality violations when compared with the expression obtained from 
perturbative factorization. Analytic expressions for the matrix elements with $1/Q^2$ precision for $1-x \gtrsim \beta/Q$ 
have also been given ($1-x \gtrsim \beta/Q$ means $m\gg 1 $, where $m$ is the 
principal quantum number of the final hadronic state, so that we can use the boundary-layer function 
to find the approximate expressions for the matrix elements). This has allowed us to obtain expressions for the moments 
$M_N$ at $O(1/Q^2)$ for finite $N$ (where we have also used the Euler-MacLaurin expansion). 
Here we have stumbled upon an unexpected result: the hadronic expression for the moments 
includes, besides the expected contributions from local matrix elements, a term at $O(1/Q^2)$ that can only be expressed 
through a non-local 4-field correlator. This non-local matrix element represents the constructive interference of two 
oscillating terms, one from the particle and the other from the antiparticle. The oscillating terms go like $(-1)^{m}$, 
and their contribution to the expansion is $o\lp \frac{1}{Q^2}\rp$, but their product doesn't oscillate and is of order $1/Q^2$. 
Upon resummation of the moments $M_N$ we have found an approximate expression for
$T$, $T^{\rm NLO}$, where $T$ is the scalar part of the tensor $T^{\mu\nu}$. This result can also be obtained using dispersion 
relations, from $\overline{W}^{\rm NLO}$, which is obtained applying the Euler-MacLaurin expansion directly to $\overline{W}$, 
the scalar part of $\overline{W}^{\mu\nu}$. $T^{\rm NLO}$ is a good approximation to $T$ for large $Q^2$ and $x_B$, 
and can be taken as a generating functional of the moments $M_N$.

We have performed the same computation of $\overline{W}$ and $T$ using perturbative factorization at one loop with $1/Q^2$ precision. 
The perturbative calculation cannot 
see the structure of the bound states, and therefore $\overline{W}_{\rm OPE}$ is a smooth function of $x$, unlike the hadronic result.
A direct comparison between hadronic and perturbative results can only be performed in the Deep Euclidean region 
through the moments $M_N$. From $\overline{W}_{\rm OPE}$ we have obtained, through a dispersion relation, 
the coefficients of the OPE of $T$, the moments $M_N^{\rm OPE}$, with $1/Q^2$ precision. We have 
checked that the OPE does get the contribution to the exact expansion from local matrix elements right, but misses the non-local one. 
This can also be seen at the level of dispersion relations, since $\overline{W}^{OPE} \not= \overline{W}^{\rm NLO}$. 
Therefore, we conclude that this expansion breaks down at NLO for DIS in the 't~Hooft model. The reason for this seems to be the 
non-analytic nature of the oscillating terms: their $(-1)^m$ behavior produces a contribution to the moments of 
$O(\beta^2/Q^2)^{1+\beta_i+\beta_j}$, which seems out of the reach of an OPE. This non-OPE structure would survive in the 
(enhanced) interference term in the form of a non-local matrix element. We have considered diagrams representing interference 
between quark and antiquark, including some that are strictly out of perturbation theory's reach, and we have seen that neither 
of them could account for the $1/Q^2$ interference term of the exact expansion, reinforcing the idea of its non-OPE nature.
The acuteness of the problem might be due to the large $N_c$ limit (therefore, one may suspect that one might run into the same 
difficulties in four-dimensional large $N_c$ QCD). Key to the appearance of the interference term is the fact that resonances 
have zero width. With finite widths the behavior of the oscillations would be milder than $(-1)^m$, which perhaps would move 
their interference down by some extra powers of $1/Q^2$. Irrespective of this last comment, and in view of the findings of this 
paper, it is evident that more work should be devoted to a more rigorous study of quark-hadron duality and OPE-violation effects 
in perturbative factorization schemes. Otherwise the errors associated to those analysis will always have a certain degree of 
uncertainty, which, at present, cannot be quantified.

The possible existence 
of OPE-breaking effects in QCD has already been discussed in the past. 
As early as in Ref. \cite{Burgio:1997hc} numerical evidence for the existence of OPE-breaking 
effects in the gluon condensate was claimed. Nevertheless, it is still 
unclear whether those effects can be associated to ultraviolet renormalons and/or higher 
orders in perturbation theory (for a recent discussion see \cite{Rakow:2005yn}). Over the years 
there has also been some discussion on the possible existence of a $\langle A^2 \rangle_{min.}$ 
condensate. This object should actually correspond to a non-local gauge-invariant condensate, though 
its explicit form is unknown for QCD \cite{Gubarev:2000eu}. Finally, there are some models that 
may produce effects that break the OPE, see for instance \cite{Dorokhov:2003kf}.
Nevertheless, those OPE-breaking effects would affect the static potential and 
the vacuum polarization. Regarding this we would like to emphasize that we do not 
find any OPE-breaking effect in the static potential or the vacuum polarization in 
the 't Hooft model. The static potential can be computed exactly in the 't~Hooft model 
within perturbation theory. Therefore, there is no room there for effects associated to a 
sort of $\langle A^2 \rangle_{min.}$ condensate. With the present precision of our 
computation, we also do not see OPE-breaking effects in the vacuum polarization \cite{Mondejar:2008dt}. 
Note that both in the case of the vacuum polarization and DIS we are talking about the same operator: 
the time-ordered product of two currents. The difference comes from the physical states 
between which we sandwiched the operator: the vacuum in the first case and one particle state in the second. 
This may point to the fact that the OPE cannot be understood as an operator equality, 
as its validity may depend on the states between which the operators are sandwiched. 


\vspace{5mm}
\noindent
{\bf Acknowledgments:}\\
This work is partially supported by the 
network Flavianet MRTN-CT-2006-035482, by the spanish 
grant FPA2007-60275, by the Spanish Consolider-Ingenio 2010
Programme CPAN (CSD2007-00042), by the catalan grant SGR2005-00916,  
and by Science and Engineering Research Canada.


\vfill
\newpage

\appendix

\section{Semileptonic B decays in the 't~Hooft model}
\label{B}
\renewcommand{\theequation}{\thesection.\arabic{equation}}
\setcounter{equation}{0}

In this appendix we present corrections to some of the formulas of Ref. \cite{Mondejar:2006ct}.
There we studied duality violations in the context of semileptonic B decays in the 't~Hooft model 
with $1/m_Q^2$ precision. The expression of the semileptonic differential decay rate was missing 
some terms, in  particular the $1/n$ terms presented in Eqs. (\ref{int1p}) and (\ref{int0p}). These 
corrections affect the computation of the moments 
but the main result remains unaltered, namely, one observes no duality violations in the 
moments with $1/m_Q^2$ precision. The reader is referred to Ref. \cite{Mondejar:2006ct} for definitions 
and a complete derivation, here we will only present the formulas and derivations that needed mending.

In Ref. \cite{Mondejar:2006ct} we computed the decay rate
\bea
\label{dGdx+}
\frac{d\Gamma^{(+)}}{dx}&=&\frac{G^2M_{H_Q}}{32\pi }\sum_{M_{m}\le M_{H_Q}}
\frac{x}{P_{H_Q}^+(1-x)}\Big| \langle cs; m|\bar{\psi}_c(0) \gamma^- Q(0)|Qs; H_Q \rangle\Big|^2
\delta\lp P_{H_Q}^--P_m^-\rp \nn\\
&=&\frac{G^2M_{H_Q}}{32\pi }\sum_{M_{m}\le M_{H_Q}}
\frac{x}{(P_{H_Q}^+)^2}\Big| \langle cs; m|\bar{\psi}_c(0) \gamma^- Q(0)|Qs; H_Q \rangle\Big|^2
\delta\lp x-1+\frac{M_m^2}{M_{H_Q}^2}\rp\nn
\,,\\
\eea

\subsubsection*{Approximate matrix elements}

\begin{figure}[ht]
\begin{center}
\includegraphics[width=0.65\columnwidth]{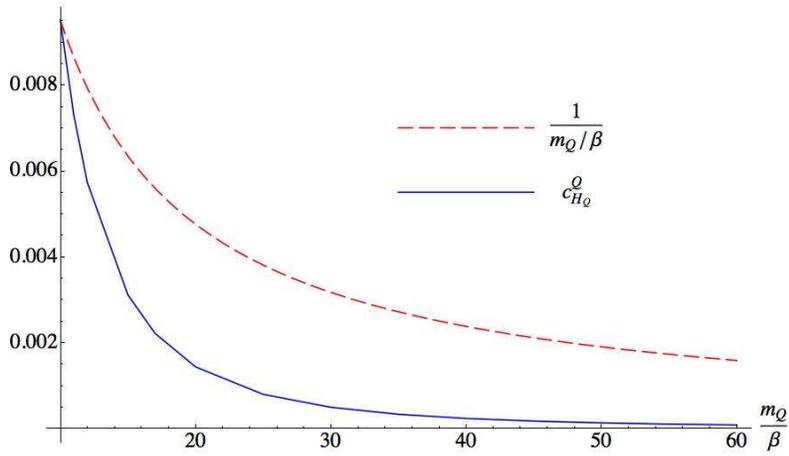}
\caption{\it Plot of the evolution of $c_{H_Q}^Q$ with $m_Q$, compared with the curve given by $1/m_Q$, normalized to match the value of $c_{H_Q}^Q$ for $m_Q=10\beta$. The mass of the spectator antiquark is $m_s=\beta$.}
\label{CQ}
\end{center}
\end{figure}

We can expand the ``diagonal" term of the matrix element of $d\Gamma^{(+)}/dx$ for large $m$ just as we did for DIS.  If we define $z\equiv \xi\beta^2/M_m^2$, the integrand will be concentrated on a small region of finite $\xi$ near the origin. For the values of $x$ such that $z(1-x)\ll x$, we can approximate
\bea
&&
\frac{1}{P_{H_Q}^+}\langle cs; m|\bar{\psi}_c \gamma^- Q|Qs; H_Q \rangle|_{diag}=\frac{m_Qm_c}{M_{H_Q}^2}\int_0^1dz\frac{\phi^{cs}_m(z)\phi^{Qs}_{H_Q}(x+(1-x)z)}{z(x+(1-x)z)}\nn\\
&&\simeq
\frac{m_Qm_c}{M_{H_Q}^2}
\left(
\frac{\phi_{H_Q}^{Qs}(x)}{x}\int_0^{1}\frac{dz}{z}{\phi}_{m}^{cs}(z)
+{\phi'}_{H_Q}^{Qs}(x)\frac{1-x}{x}\int_0^{1}dz {\phi}_{m}^{cs}(z)
-\frac{\phi_{H_Q}^{Qs}(x)}{x^2}(1-x)\int_0^{1}dz{\phi}_{m}^{cs}(z)
\right)
\nn
\\
\nn
&&=
\pi\beta\frac{m_Q}{M_{H_Q}^2 x}
\left[ \lp 1+\frac{m_{c,R}^2+m_{s,R}^2}{2M_m^2}+\frac{m_cm_s}{M_m^2}(-1)^m-\frac{1-x}{x}\frac{m_c^2+(-1)^mm_cm_s}{M_m^2}\rp
\phi^{Qs}_{H_Q}(x)\right.\nn\\
&&\left.+(1-x)\frac{(m_c^2+(-1)^mm_cm_s)}{M_m^2}{\phi'}^{Qs}_{H_Q}(x)+o\left(\frac{1}{m_Q^2}\right)\right]
\ , \nn
\eea
\be
\label{eqlayerHQ}
\ee
where we make the counting $M_m^2 \sim m_Q^2$. 
Note that in the last equality we could use $M_m^2= M_{H_Q}(1-x)$, since the physical matrix element is only defined for the values 
of $x$ given by the delta of momentum conservation,
\be
M_m^2=M_{H_Q}^2(1-x_m) \ ,
\ee
and/or use $M_m^2 \simeq m\pi^2\beta^2$, since the above computation is meant for large values of $m$. 
Unlike in DIS, however, the condition $z(1-x)\ll x$ does not always hold for physical values of $x$: in this case we have
\be
z\frac{1-x_m}{x_m}\to \xi\frac{\beta^2}{M_m^2}\frac{1-x_m}{x_m}=\xi\frac{\beta^2}{M_{H_Q}^2x_m} \ .
\ee
Thus, for $x_m\sim \beta^2/M_{H_Q}^2$, which corresponds to the largest excitations available for the decay, our approximation in Eq. (\ref{eqlayerHQ}) does not hold. In this region we should approximate the ``diagonal" matrix element by
\bea
\nn
\frac{1}{P_{H_Q}^+}\langle cs; m|\bar{\psi}_c \gamma^- Q|Qs; H_Q \rangle|_{diag,\,x_m\sim\beta^2/M_Q^2}
&\simeq& 
\frac{m_Qm_c}{M_{H_Q}^2}\int_0^1dz\frac{\phi^{cs}_m(z)}{z}c_{H_Q}^Q(x_m+z)^{\beta_Q-1}
\\
&\simeq& \frac{m_Q\pi\beta c_{H_Q}^Q}{M_{H_Q}^2}\ .
\label{eqborder}
\eea
The contribution from this region is suppressed by the factor of $c_{H_Q}^Q$, which goes to zero as 
$m_Q\to \infty$ faster than $1/m_Q$, as Fig. \ref{CQ} shows. Therefore, it can be neglected. 

The ``off-diagonal" matrix element remains as it is shown in Ref. \cite{Mondejar:2006ct}. Overall, we write the total=``diagonal"+``off-diagonal" matrix element in the following way for large $m$ and $m_Q$ (up to a global sign, and setting $x=x_m$):
\bea
\label{optionBR}
&&\int_0^1dz \phi^{cs}_m(z)\phi^{Qs}_{H_Q}(z)\simeq\pi\beta\frac{m_{Q,R}}{M_{H_Q}^2 x_m}\nn \\
&&
\times\left[ \lp 1+\frac{m_{c,R}^2+m_{s,R}^2}{2m\pi^2\beta^2}+\frac{m_cm_s}{m\pi^2\beta^2}(-1)^m-\frac{m_{c,R}^2+(-1)^mm_cm_s}{m_Q^2 x_m}\rp
\phi^{Qs}_{H_Q}(x_m)\right.\nn\\
&&\left.+\frac{(m_{c,R}^2+(-1)^mm_cm_s)}{m_Q^2}{\phi'}^{Qs}_{H_Q}(x_m)+o\left(\frac{1}{m_Q^2}\right)\right]
\ . 
\eea

We will need as well an approximate expression for the matrix element in the limit\newline $1-x_m\lesssim\beta^2/m_{Q}^2$ (corresponding to the lowest resonances of the meson $c\bar{s}$). Following the same procedure that led to Eq. (\ref{1x}) in the case of DIS, we find
\bea
\label{optionBS}
&&\int_0^1dz \phi^{cs}_m(z)\phi^{Qs}_{H_Q}(z)|_{x_m=1-M_m^2/M_{H_Q}^2,\, M_{H_Q}^2\to\infty}=\frac{m_Qm_c}{M_{H_Q}^2}c^s_{H_Q}\lp \frac{M_m^2}{M_{H_Q}^2}\rp^{\beta_s}\int_0^1dz\frac{\phi_m^{cs}(z)}{z}(1-z)^{\beta_s}\nn\\
&&+\frac{m_Qm_c}{M_{H_Q}^2}c^s_{H_Q}\lp \frac{M_m^2}{M_{H_Q}^2}\rp^{1+\beta_s}\int_0^1dz\frac{\phi_m^{cs}(z)}{z}(1-z)^{1+\beta_s}-\frac{\beta^2}{M_{H_Q}^2}\lp\frac{M_m^2}{M_{H_Q}^2}\rp^{\beta_s}c^s_{H_Q}\sum_{n'=0}^{\infty}(-1)^{n'} \nn\\
&&\times\int_0^{1/(1-x)}du\int_0^1dv\int_0^1dz\frac{\phi_m^{cs}(z)\phi_{n'}^{Qc}(1-u(1-x))\phi_{n'}^{Qc}(v)}{(u+z)^2}\lp(1-z)^{\beta_s}-(1+u)^{\beta^s
}\rp \nn\ .\\
\eea
This contribution is suppressed by a factor $1/M_{H_Q}^{2\beta_s}$ with respect to the leading term in Eq. (\ref{optionBR}), but it is enhanced with respect to the $1/m_Q^2$ corrections, just like in DIS.
\bigskip

For ``intermediate" values of  $x$ (away from the boundaries $x\sim \beta^2/m_Q^2$ and $1-x\sim \beta^2/m_Q^2$) the differential decay rate reads then, at $O(1/m_Q^2)$,
\bea
\label{dgammalayer}
&&
\frac{d\Gamma^{(+)}}{dx}=\frac{1}{2}\sum_{M_m \leq M_{H_Q}} 
\frac{G^2M_{H_Q}}{4\pi}
\frac{m_{Q,R}^2}{M^2_{H_Q}}
\frac{\pi^2 \beta^2}{M^2_{H_Q}} 
\frac{1}{x}\phi^{Qs}_{H_Q}(x) \nn\\
&&\times\left[ \lp 1+\frac{m_{c,R}^2+m_{s,R}^2}{m\pi^2\beta^2}+2\frac{m_cm_s}{m\pi^2\beta^2}(-1)^m-2\frac{m_{c,R}^2+(-1)^mm_cm_s}{m_Q^2 x}\rp
\phi^{Qs}_{H_Q}(x)\right.\nn\\
&&\left.+2\frac{(m_{c,R}^2+(-1)^mm_cm_s)}{m_Q^2}{\phi'}^{Qs}_{H_Q}(x)\right]\delta\lp x-1+\frac{M_m^2}{M_{H_Q}^2}\rp\ . 
\eea

\subsubsection*{Moments}

The differential decay rate is not a very well defined object in the large $N_c$, since it 
becomes either infinity or zero. Its comparison with the expressions 
obtained from effective theories that use perturbative factorization is 
not possible, as they yield a smooth function of $x$, so we turn to moments, which we define as:
\be
\label{momdefB}
M_N \equiv \int_{x_{min.}}^{x_{max.}} dx x^{N-1} \frac{d\Gamma}{dx}
\,,
\ee
where $x_{min.}$ and $x_{max.}$ are given by
\be
x_{min.}=1-\frac{M_{m^*}}{M_{H_Q}^2}\ ,\quad\quad x_{max.}=1-\frac{M_{0}^2}{M_{H_Q}^2} \ ,
\ee
where $M_{m^*}$ is the mass of the maximum resonance of the meson $c\bar{s}$ allowed by momentum conservation, and $M_{0}$ 
is the mass of its ground state.

The calculation of the moments using Eq. (\ref{dgammalayer}) goes along the similar lines than our previous calculation of 
moments in DIS. As the differential decay rate is a series of deltas, the moments will be a sum over an index $m$ that, as in DIS, we will 
rewrite using the Euler-Maclaurin expansion shown in Eq. (\ref{EulerMacLaurin}),
where now 
the limits $m=m^*$ and $m=0$  correspond to $x=x_{min.}$ and $x=x_{max.}$, respectively. 
At the limit $x=x_{min.}$ ($m=m^*$) we can use the matrix element given in Eq. (\ref{eqborder}). The contributions from both $f(m^*)$ and $f^{(n)}(m^*)$ will be suppressed by a relative factor of $1/m_Q^2(c_{H_Q}^Q)^2<1/m_Q^4$ with respect to the leading term; therefore, we neglect them. At $x=x_{max.}$ ($m=0$) we use Eq. (\ref{optionBS}) to represent the matrix element, and so we see that both $f(0)$ and $f^{(n)}(0)$ are suppressed by a relative factor of $O((\beta^2/m_{Q^2})^{1+2\beta_s})$. The integral goes from $m=0$ to $m=m^*$ (from $x=x_{min.}$ to $x=x_{max.}$), so in principle we should divide it into three pieces: one for low values of $x$, in which we use Eq. (\ref{eqborder}) for the matrix element; another for intermediate values of $x$, in which we use Eq. (\ref{optionBR}); and another for high values of $x$, in which we use Eq. (\ref{optionBS}). However, if we just insert Eq. (\ref{optionBR}) in the lower boundary, we will be making an error of $o(1/m_Q^2)$, and the contribution from the higher boundary is $O((\beta^2/m_Q^2)^{1+2\beta_s})$ whether we use Eq. (\ref{optionBR}) or Eq. (\ref{optionBS}).

Summing up, we take Eq. (\ref{dgammalayer}), make the change $\sum_m\to\int dm$, and insert it into Eq. (\ref{momdefB}), and what we obtain is (using Eq. (\ref{massn2}) for the spectrum)
\bea
\label{MNlayersmeared}
M_N&\simeq& \frac{G^2M_{H_Q}}{4\pi}
\frac{m_{Q,R}^2}{M^2_{H_Q}}\int_{x_{min.}}^{x_{max.}} dx \,x^N\frac{1}{x^2}\phi^{Qs}_{H_Q}(x)\left[ \lp 1-2\frac{m_{c,R}^2}{m_{Q}^2 x}\rp
\phi^{Qs}_{H_Q}(x)+2\frac{m_{c,R}^2}{m_{Q}^2}{\phi'}^{Qs}_{H_Q}(x)\right] \nn\\
&\simeq&\frac{G^2M_{H_Q}}{4\pi}
\frac{m_{Q,R}^2}{M^2_{H_Q}}
\int_{x_{min.}}^{x_{max.}}dx\, x^{N}
\frac{\left[\phi_{H_Q}^{Qs}\left(x+\frac{m_{c,R}^2}{m_{Q,R}^2}\right)\right]^2}{\lp x+\frac{m_{c,R}^2}{m_{Q,R}^2}\rp^2}\nn\\
&=&\frac{G^2M_{H_Q}}{4\pi}\frac{m_{Q,R}^2}{M_{H_Q}^2}
\int_{x_{min.}+\frac{m_{c,R}^2}{m_{Q,R}^2}}^{x_{\max.}+\frac{m_{c,R}^2}{m_{Q,R}^2}} dx
\lp1-\frac{m_{c,R}^2}{xm_{Q,R}^2}\rp^N  
 x^{N}\frac{[\phi_{H_Q}^{Qs}\lp x\rp]^2}{x^2} \nn\\
&\simeq&
\frac{G^2M_{H_Q}}{4\pi}\frac{m_{Q,R}^2}{M_{H_Q}^2}
\lp 1-\frac{m_{c,R}^2}{m_{Q,R}^2}\rp^N 
\int_0^1 
 \frac{dx}{x^2} x^{N}[\phi_{H_Q}^{Qs}\lp x\rp]^2
 \,.
\eea
There has been a number of approximations here. In the first line we neglected the oscillating $(-1)^m$ terms, 
since their contribution is suppressed by a relative factor of $o(\beta^2/m_Q^2)$. In the second line we 
have reshuffled the NLO correction in a way that is correct at the accuracy of the calculation. 
And in the last line we have used $\frac{m_{c,R}^2}{xm_{Q,R}^2} \simeq \frac{m_{c,R}^2}{m_{Q,R}^2}$, 
which is correct again with the accuracy of our calculation (this approximation is wrong as $x\to 0$, 
but there the asymptotic behavior of $\phi_{H_Q}^{Qs}(x)$ ensures that the error be of $o(1/m_Q^2)$), 
and we have also extended the lower limit of integration from $x=x_{min.}+\frac{m_{c,R}^2}{m_{Q,R}^2}$ 
to $x=0$, and the upper limit from $x=x_{max.}+\frac{m_{c,R}^2}{m_{Q,R}^2}$ to $x=1$; in the lower limit 
the error will be $o(1/m_Q^2)$, and in the upper limit it will be of $O((\beta^2/m_Q^2)^{1+2\beta_s})$.

The right-hand side of Eq. (\ref{MNlayersmeared}) contains some implicit dependence on the heavy quark mass, since so far we have used the exact $H_Q$-meson. If we perform 
an explicit expansion in $1/m_Q$, one obtains for the first moments, up to ${O}(1/m_Q^3)$,
\be
M_0=\frac{G^2m_{Q}}{4\pi}
\left[1+\frac{\langle t \rangle}{m_Q}-\frac{\langle t\rangle^2- \langle t^2\rangle+\beta^2}{2m_Q^2}+{O}\left(\frac{1}{m_Q^3}\right)\right]
\,,
\ee
\be
M_1=
\frac{G^2m_{Q}}{4\pi}
\left[1+\frac{ \langle t\rangle^2- \langle t^2\rangle+\beta^2-2m_c^2}{2m_Q^2}+{O}\left(\frac{1}{m_Q^3}\right)\right]
\,,
\ee 
\be
M_2=
\frac{G^2m_{Q}}{4\pi}
\left[1-\frac{\langle t\rangle}{m_Q}+\frac{3\langle t\rangle^2- \langle t^2\rangle+3\beta^2-4m_c^2}{2m_Q^2}+{O}\left(\frac{1}{m_Q^3}\right)\right]
\,,
\ee
where the static limit expectation values are defined in Ref. \cite{Mondejar:2006ct}.

Eq. (\ref{MNlayersmeared}) is exactly the result we showed in Ref. \cite{Mondejar:2006ct}. Therefore, the conclusion we reached there for the moments still holds: they show no duality violations with $1/m_Q^2$ precision. 

\vfill
\newpage


\end{document}